\documentclass[letter,11pt]{article}
\usepackage[top=1in,bottom=1in,left=1in,right=1in]{geometry}
\usepackage{slashed}
\usepackage{epsfig}
\usepackage{comment}
\usepackage{cancel}
\usepackage{bbm}
\usepackage{array}
\usepackage{bigints}
\usepackage{booktabs}
\usepackage{xcolor}
\usepackage{dsfont}
\usepackage{float}
\usepackage{framed}
\usepackage{graphicx}
\usepackage{indentfirst}
\usepackage{mathrsfs}
\usepackage{multirow}
\usepackage{subdepth}
\usepackage{titlesec}
\usepackage[dotinlabels]{titletoc}
\usepackage{wrapfig}
\usepackage[all]{xy}
\usepackage[vcentermath]{youngtab}
\usepackage{relsize}
\usepackage{hyperref}
\numberwithin{equation}{section}
\usepackage{soul}

\usepackage[utf8]{inputenc}
\usepackage{slashed}
\usepackage{amsmath}
\usepackage{amsfonts}
\usepackage{amssymb}
\usepackage{cite}
\usepackage{mathtools}

\usepackage{cleveref}
\crefname{section}{§}{§§}
\Crefname{section}{§}{§§}

 \def\p{\partial}
 \def\bz{{\bar z}}

\def\0{{(0)}}
\def\1{{(1)}}
\def\2{{(2)}}

\def\<{\langle }
\def\>{\rangle }

\newcommand{\bea}{\begin{eqnarray}}
\newcommand{\eea}{\end{eqnarray}}
\newcommand{\be}{\begin{equation}}
\newcommand{\ee}{\end{equation}}
\newcommand{\ba}{\begin{align}}
\newcommand{\ea}{\end{align}}

   \makeatletter
  \let\over=\@@over \let\overwithdelims=\@@overwithdelims
  \let\atop=\@@atop \let\atopwithdelims=\@@atopwithdelims
  \let\above=\@@above \let\abovewithdelims=\@@abovewithdelims
\renewcommand\section{\@startsection {section}{1}{\z@}%
                                   {-3.5ex \@plus -1ex \@minus -.2ex}
                                   {2.3ex \@plus.2ex}%
                                   {\normalfont\large\bfseries}}

\renewcommand\subsection{\@startsection{subsection}{2}{\z@}%
                                     {-3.25ex\@plus -1ex \@minus -.2ex}%
                                     {1.5ex \@plus .2ex}%
                                     {\normalfont\bfseries}}

\newcommand{\beq}{\begin{equation}}
\newcommand{\eeq}{\end{equation}}
\newcommand{\beqa}{\begin{eqnarray}}
\newcommand{\eeqa}{\end{eqnarray}}
\newcommand{\beqar}{\begin{eqnarray*}}

\newcommand{\ve}{{\varepsilon}}

\def\[{\big[}
\def\]{\big]}

\def\ve{{\varepsilon}}

\def\bz{{\bar z}}
\def\ba{{\bar a}}

\def\be{{\bar \epsilon}}

\def\eps{\epsilon}

\def\+{{(+)}}
\def\-{{(-)}}
\def\0{{(0)}}
\def\1{{(1)}}
\def\2{{(2)}}
\def\3{{(3)}}

\def\be{\begin{equation}}
\def\ee{\end{equation}}

\begin{document}
\begin{titlepage}
\unitlength = 1mm
\ \\
\vskip 3cm
\begin{center}

{\huge{\textsc{${\rm w}_{1+\infty}$ in 4D Gravitational Scattering}}}

\vspace{1.25cm}
Elizabeth Himwich$^{*}$ and Monica Pate$^{ \dagger}$

\vspace{.5cm}

$^*${\it  Princeton Center for Theoretical Science, Princeton University, Princeton, NJ 08544}\\ 
$^\dagger${\it   The Center for Cosmology and Particle Physics, New York University, New York, NY 10003}\\

\vspace{0.8cm}

\begin{abstract} 

In four-dimensional asymptotically flat spacetimes, an infinite tower of soft graviton modes is known to generate the symmetry algebra of ${\rm w}_{1+\infty}$ at tree-level. Here we demonstrate that the symmetry action follows from soft graviton theorems and acts non-trivially on massive scalar particles. By generalizing previous analyses that were specifically tailored to the scattering of massless particles, our results clarify that ${\rm w}_{1+\infty}$ symmetry is a universal feature of tree-level gravitational scattering in four-dimensional asymptotically flat spacetimes and originates from minimally-coupled gravitational interactions.  In addition, we show that the ${\rm w}_{1+\infty}$ symmetry acts non-diagonally on massive states by mixing an infinite number of conformal families. We also present a concrete example of non-local behavior on the celestial sphere in the presence of massive scattering states.

\end{abstract}

\vspace{1.0cm}

\end{center}

\end{titlepage}

\pagestyle{empty}
\pagestyle{plain}

\def\vx{{\vec x}}
\def\p{\partial}
\def\po{$\cal P_O$}
\def\i{{\rm initial}}
\def\f{{\rm final}}

\pagenumbering{arabic}
 

\tableofcontents

\section{Introduction}

Quantum field theory has proven to be a remarkably successful framework for simplifying and unifying fundamental particle physics. Nevertheless, because gravitational quantum field theories are non-renormalizable, they ostensibly do not predict a unique theory of quantum gravity. Holography, a proposed correspondence between theories of quantum gravity and non-gravitational quantum field theories in fewer spacetime dimensions, offers an alternative way to constrain quantum gravity using quantum field theory.  The holographic correspondence thus reframes quantum field theory as the central tool for describing microscopic physics, including in the context of quantum gravity. 

In the decades following the discovery of the AdS/CFT correspondence \cite{Maldacena:1997re,Witten:1998qj}, holography has been put to the test both within AdS/CFT \cite{Heemskerk:2009pn,Maldacena:2015iua,Caron-Huot:2021enk} and by a number of new holographic proposals \cite{Strominger:2001pn,deBoer:2003vf,Guica:2008mu,Costello:2018zrm}.  In recent years, celestial holography has emerged as an important new type of holography. Celestial holography posits a duality between scattering in asymptotically flat four-dimensional (4D) spacetime and a two-dimensional conformal field theory (2D CFT) on the celestial sphere. Reviews and references on the subject can be found in \cite{Pasterski:2021rjz,Raclariu:2021zjz,Pasterski:2021raf}. The celestial duality is motivated by the isomorphism between the Lorentz group in four dimensions ${\rm SO}(1,3)$ and the global conformal group in two dimensions ${\rm SL} (2, \mathbb{C})$.  This isomorphism ensures that scattering amplitudes for particles in highest weight representations of ${\rm SL}(2, \mathbb{C})$ transform under Lorentz transformations like correlation functions of primary operators in a 2D CFT under global conformal transformations \cite{deBoer:2003vf,Kapec:2016jld,Cheung:2016iub,Pasterski:2016qvg,Pasterski:2017kqt}. The basis of states in highest weight representations of ${\rm SL}(2, \mathbb{C})$ is referred to as the conformal primary basis and scattering amplitudes in this basis are known as celestial amplitudes. 

Holographic proposals stipulate that the defining properties of a quantum field theory are imprinted in a corresponding dual gravitational theory, and vice versa. However, identifying these signatures appropriately can be a difficult task. Locality, in particular, is a key property of quantum field theory with a nontrivial holographic imprint even in the well-studied example of AdS/CFT. In celestial holography, the full implications of locality of the dual 2D CFT are yet to be determined. 2D locality may turn out to require properties of celestial amplitudes that appear nontrivial or are obscure from the perspective of scattering in 4D.
    
A simple notion of 2D locality does readily appear in the 4D scattering of massless particles.  Two essential ingredients underlie this result. First, the wavefunction of a massless particle in a momentum eigenstate localizes to a point on a spatial cross section of null infinity that can be identified with the spatial direction of the momentum four-vector. Second, a state in a conformal primary basis for massless particles can be constructed from the subset of momentum eigenstates with momenta pointing in a fixed direction.  The resulting state transforms like a primary operator located at the associated point on a spatial cross section of null infinity. In this construction, the location of operators in 2D thus coincides with a projective location of particles in 4D. Consequently, 2D local behavior, such as singularities when operators approach one another, has a simply origin and interpretation in 4D physics. 
    
On the other hand, a notion of 2D locality for massive particles is less straightforward.  In this case, the wavefunctions of momentum eigenstates localize to points on the resolution of timelike infinity to a spacelike hyperboloid (Euclidean ${\rm AdS}_3$) \cite{Campiglia:2015qka,Campiglia:2015kxa}. States in a conformal primary basis have been constructed by convolving momentum eigenstates with ${\rm AdS}_{3}/{\rm CFT}_2$ bulk-to-boundary propagators and therefore receive contributions from all momenta, not just a momentum subset  \cite{Pasterski:2016qvg,Pasterski:2017kqt}. This can be understood geometrically as the statement that the subset of ${\rm SL}(2, \mathbb{C})$ transformations that preserve a point on the boundary of ${\rm AdS}_3$ preserve no proper subset of points in the bulk of ${\rm AdS}_3$. A derivation of this statement is provided in Appendix \ref{app_b}. As a result, for massive particles, there is no  direct analog of the massless correspondence between 2D and 4D locality.\footnote{A similar conclusion seems to apply to the conformal primary of basis of massless particles whose construction involves an additional shadow transformation.  It would very interesting, but beyond the scope of this paper, to investigate whether locality is restored when working with a discrete basis for massive particles analogous to the massless basis in \cite{Freidel:2022skz,Cotler:2023qwh}.} 
    
While such a correspondence is not necessary for 2D locality to exist -- indeed, in AdS/CFT the regimes of bulk and boundary locality do not coincide  -- its absence nevertheless calls into question the realization of 2D locality for massive particles.  So far, there are very few examples of explicit formulas for celestial amplitudes containing massive external particles.\footnote{This is especially true beyond three-point correlators, where the singularity structure is no longer tightly constrained by 2D global conformal symmetry. \cite{Casali:2022fro,Banerjee:2023rni,Taylor:2023bzj} consider amplitudes with a massive source that breaks translation invariance. \cite{Ball:2023ukj} considers a $\Delta=0$ massive celestial scalar coupled to gravitons. Several works have considered partial wave decompositions and factorization of four-point celestial amplitudes with an internal massive exchange \cite{Cardona:2017keg,Lam:2017ofc,Nandan:2019jas,Chang:2021wvv,Chang:2023ttm}.} The possibility therefore remains that local behavior emerges only when computing a full celestial amplitude.
    
In the hope of inspiring more rigorous studies of 2D locality, we now present an explicit example to illustrate that a local 2D description of massive particles must involve modifications to the current prescription.  Consider the scattering of a conformally soft photon $J$, two massless scalars $\phi$ of charge $Q$, and a massive scalar $\Phi$ of charge $-2Q$.  The celestial amplitude for this process, which to our knowledge has not yet appeared in the literature and we derive in Appendix \ref{four-point-section}, is given by 
\begin{equation} \label{non-local-amp}
    \begin{split} 
        \langle \phi_{\Delta_1}& (z_1, \bar{z}_1) J(z_2, \bar{z}_2)
            \phi_{\Delta_3} (z_3,\bar{z}_3) \Phi_{\Delta_4} (z_4, \bar{z}_4) \rangle\\         
        & =Q  \left[ \frac{1}{z_{21}}+ \frac{1}{z_{23}} - \frac{2}{z_{24}}\left(1+\frac{z_{34}}{z_{23}} \frac{ z-1 +\bar{z} \, {}_2F_1(1,\frac{\Delta_1-\Delta_3+\Delta_4}{2},\Delta_4;1- \frac{z}{z-1}\frac{\bar{z}}{\bar{z}-1})  }{z - 1 + \bar{z} } \right) \right] 
        \\& \quad\quad \quad  \times  \langle \phi_{\Delta_1} (z_1,\bar{z}_1)
            \phi_{\Delta_3} (z_3,\bar{z}_3) \Phi_{\Delta_4} (z_4, \bar{z}_4) \rangle,
    \end{split}
\end{equation} 
where the three-point celestial amplitude is 
\begin{equation}
    \begin{split}
        \langle \phi_{\Delta_1} (z_1,&\bar{z}_1)\phi_{\Delta_3} (z_3,\bar{z}_3)
            \Phi_{\Delta_4} (z_4, \bar{z}_4) \rangle\\
         &= \frac{g}{8}\left(\frac{m}{2} \right)^{\Delta_1+\Delta_3-4} \frac{B \left(\frac{\Delta_4 + \Delta_1-\Delta_3}{2},\frac{\Delta_4 - \Delta_1+\Delta_3}{2}\right) }{(z_{41} \bz_{41})^{\frac{1}{2}(\Delta_4+\Delta_1-\Delta_3)}
		(z_{13} \bz_{13})^{\frac{1}{2}(\Delta_1+\Delta_3-\Delta_4)}
		(z_{34} \bz_{34})^{\frac{1}{2}(\Delta_3+\Delta_4-\Delta_1)} }
    \end{split}
\end{equation}
and $z$ is the conformal cross ratio
\begin{equation}
    z = \frac{z_{12}z_{34}}{z_{13}z_{24}}.
\end{equation}
The asymmetry in scalars 1 and 3 results from the assumption that ${\rm Re}(\Delta_4) > {\rm Re}(\frac{\Delta_1-\Delta_3+\Delta_4}{2})>0$, which is needed for the integral expression \eqref{hypergeo-int} for the hypergeometric function to converge.  Appendix \ref{four-point-section} includes a full derivation of this formula, as well as more detailed explanations of the statements that follow. 

The hypergeometric function in the amplitude \eqref{non-local-amp} presents a striking difference from the corresponding all-massless celestial amplitude.  Momentarily ignoring the additional complexity of this hypergeometric function, note that the first two terms contain simple poles in $z_2$ at $z_1$ and $z_3$ with residues $Q_1 = Q$ and $Q_3 = Q$, respectively. These two terms, along with only the first contribution in the round parentheses (i.e.~the term $-\frac{2}{z_{24}}$), would be the complete expression in a standard local 2D conformal field theory, which would have a third simple pole at $z_2 \to z_4$ with residue $Q_4 = -2Q$.  By contrast, in \eqref{non-local-amp} we find an additional contribution given by the second term in round parentheses. Together, the third term (including both contributions in round parentheses) naively appears to contain simple poles in $z_2$ at $z_3$, $z_4$, and $z_{\star}$, where $z_{\star}$ is the value of $z_2$ for which $z+\bar{z}-1$ vanishes.  However, a more careful analysis of residues reveals that the third term actually contains no simple poles. The celestial amplitude therefore only contains simple poles at the locations of  each of the massless particles (from the first two terms, at $z_1$ and $z_3$) but not the massive particle ($z_4$).  To summarize, the simple poles in this celestial amplitude differ from those in a traditional conformal field theory with a $U(1)$ current by the absence  of a pole at the location of the charged massive particle. Still, the residues at $z_1$ and $z_3$ are each $Q$, as in a standard conformal field theory. 

To reconcile the discrepancy between the singularity structure of the celestial amplitude and that of standard CFT correlation functions, recall that the $(1,0)$ conformal weight carried by $J$ guarantees that it must respect a conservation law of the form 
\begin{equation} \label{current_conservation}
    \begin{split}
        \oint_{\infty}  dz~  \langle J(z) \mathcal{O}_1(z_1, \bar{z}_1)
             \cdots \mathcal{O}_n(z_n, \bar{z}_n) \rangle = 0.
    \end{split}
\end{equation}
Since the sum of the residues of the simple poles in the celestial amplitude \eqref{non-local-amp} is $2Q \neq 0$, this implies that there must be some other singularity in the amplitude that cancels the simple pole contribution.  Indeed, the hypergeometric function ${}_2F_1(a,b,c;x)$ contains a branch cut in $x$ extending from $x = 1$ to $x =\infty$, which we choose to extend along the positive real axis.  In the $z_2$ plane, this corresponds to a branch cut extending between the locations of the massless particle insertions, $z_1$ to $z_3$.  In Appendix \ref{four-point-section}, we show that this branch cut accounts for precisely the missing contributions that allow \eqref{current_conservation} to hold. 

The example \eqref{non-local-amp} is especially instructive because it is a particularly clean and robust result, for the following reasons. First, all of the momentum-conserving delta functions are absorbed by the integral transformations to the conformal primary basis. The resulting celestial amplitudes are consequently smooth in the sense that they do not contain delta-function singularities, unlike their massless counterparts.  Second, \eqref{non-local-amp} is essentially fixed by kinematics and thus unaffected by loop corrections. In particular, neither the leading soft photon theorem \cite{Weinberg:1965nx} nor the kinematic dependence of three-point amplitudes receive loop corrections \cite{Arkani-Hamed:2017jhn}. Furthermore, as detailed in Appendix \ref{four-point-section}, the contour arguments and other components of the analysis treat $z$ and $\bar{z}$ as independent complex variables, and so it is not obvious how a different choice of spacetime signature could resolve this non-locality.  

Massive particles are a definitive feature of our universe, and more generally are often instrumental to the consistency of theories of quantum gravity, including celestial holography. For example, in the context of celestial holography, an exchange of massive particle states was found to appear in the conformal block decomposition of a four-point celestial amplitude with only massless external particles \cite{Atanasov:2021cje}. This result is likely quite general for the following reason. Celestial holography is a manifestly conformally  covariant framework that naturally involves physics at all energies and therefore 
may require regulators. Because the number of dimensions remains fixed, it would not be surprising if all manifestly conformally covariant regulators are of ``Pauli-Villars" form, i.e.~involve the addition of massive particles. Outside of celestial holography, massive particles are not only ubiquitous in string theory, our known example of a consistent theory of quantum gravity, but also frequently feature in general consistency conditions for quantum gravity \cite{Camanho:2014apa,Caron-Huot:2016icg,Kologlu:2019bco,Bern:2021ppb,Caron-Huot:2022ugt,Caron-Huot:2022jli,Haring:2023zwu}.

This compelling evidence from both theory and observation suggests that massive particles should be included in celestial holography, although their celestial amplitudes appear to admit nonlocal behavior. Therefore -- especially without 2D locality -- it is important to establish general properties of massive states that control their dynamics.  Symmetries naturally supply such dynamical constraints and are the subject of our work.\footnote{Indeed, celestial amplitudes are known to admit other forms of non-local behavior, but some of these instances are regarded with less concern precisely because they are understood as result of symmetry.  Specifically, four-point celestial amplitudes for massless particles are known to admit a delta-function singularity constraining the conformal cross ratio to be real.  However, this constraint can be identified as a remnant of 4D translation symmetry. Additional forms of non-locality are discussed in \cite{Ball:2022bgg,Ball:2023sdz}. }  In particular, we establish that massive particles transform non-trivially under action of the ${\rm w}_{1+\infty}$ symmetry generated by an infinite tower of soft gravitons.
   
The existence of a ${\rm w}_{1+\infty}$ symmetry in celestial holography was first discovered in \cite{Guevara:2021abz,Strominger:2021lvk} by analyzing the algebra of an infinite collection of soft graviton currents that are universal to any celestial holographic dual.  The non-trivial action of ${\rm w}_{1+\infty}$ on generic massless particles was subsequently determined in the conformal primary basis in \cite{Himwich:2021dau}.  Both analyses rely heavily on the 2D local behavior of massless particles described above as well as a conformally-invariant notion of left and right-moving coordinates that permits an asymmetric treatment of $z$ and $\bar{z}$.\footnote{Recently, this asymmetry has been elevated in celestial holography to the extent that the left-moving coordinate $z$ is treated as a point in the 2D spacetime, while the anti-holomorphic coordinate $\bar{z}$ is treated  as an auxiliary parameter, akin to a structure constant \cite{Costello:2022wso,Monteiro:2022lwm,Guevara:2022qnm}. Several of these works were inspired by investigations in twistor space, where this perspective arises naturally \cite{Witten:2003nn,Arkani-Hamed:2009hub}.  \cite{Sharma:2021gcz} first established the connection of twistor space to celestial holography. } By contrast, as mentioned earlier in the introduction, massive particles do not obviously enjoy a notion of 2D locality.  Likewise, there is no conformally-invariant notion of left and right-moving coordinates in the bulk of ${\rm AdS}_3$ and consequently the massive analysis demands a uniform treatment of all coordinates. While these may appear to be mere technical differences between massive and massless particles, they inhibit a simple generalization of the massless analysis to the massive case. Indeed, while the number of investigations regarding ${\rm w}_{1+\infty}$ in celestial holography has risen rapidly \cite{Himwich:2021dau,Jiang:2021ovh,Jiang:2021csc,Adamo:2021lrv,Ahn:2021erj,Ball:2021tmb,Mago:2021wje,Freidel:2021ytz,Ahn:2022vfw,Ahn:2022qex,Ahn:2022orj,Ren:2022sws,Bu:2022iak,Costello:2022wso,Monteiro:2022lwm,Guevara:2022qnm,Monteiro:2022xwq,Mason:2022hly,Banerjee:2023zip,Bittleston:2023bzp, Drozdov:2023qoy,Saha:2023abr,Ahn:2023mdg,Banerjee:2023jne,Taylor:2023ajd}, none yet pertain directly to massive particles.\footnote{There is a construction \cite{Crawley:2023brz} of a massive self-dual black hole state with a tower of $\rm{w}_{1+\infty}$ charges, although this state is built as a coherent sum of massless graviton celestial operators and relies on the asymmetry in $z$ and $\bar{z}$ in the self-dual  context.} 

In this work, we generalize  the analysis of \cite{Himwich:2021dau} to incorporate massive particles and thereby establish that the symmetry acts non-trivially on massive particles. The generalization involves modifying the formulas for the infinite tower of momentum-space soft theorems found in \cite{Hamada:2018vrw,Li:2018gnc} to facilitate a simple transformation to the conformal primary basis.  In particular, we demand that the associated soft factors transform like conformal primaries under ${\rm SL} (2, \mathbb{C})$.  We find that these improved soft factors admit descendants that also transform like primaries (i.e.~primary-descendants).  Our analysis thereby affirms that the shortened conformal multiplet structure found in \cite{Guevara:2021abz,Pasterski:2021dqe,Pasterski:2021fjn} is unaffected by the specific particle content of the theory (i.e.~massive versus massless) and thus is an intrinsic of property of the soft gravitons and their ${\rm SL} (2, \mathbb{C})$ representations.  We then find that the symmetry action on momentum-space massive scalar particles takes a particularly simple form, which we use to establish that it respects the algebra of ${\rm w}_{1+\infty}$.  Finally, we transform our results to the conformal primary basis and recover the known action of Poincar\'e generators on massive scalar particles in  \cite{Stieberger:2018onx,Law:2019glh}. We discover that the additional ${\rm w}_{1+\infty}$ generators beyond those that generate the Poincar\'e algebra act in a highly non-diagonal manner on massive particles in the conformal primary basis.  Specifically, they transform a state of definite conformal weight into a combination of a infinite number of states with conformal weights differing by integer values from that of the original state.

The paper is organized as follows. First, we present our conventions and carefully review the ${\rm SL}(2, \mathbb{C})$ transformation properties of celestial amplitudes in Section \ref{sec:review}.   In Section \ref{sec:softw}, we derive the action of ${\rm w}_{1+\infty}$ on hard particles directly from the infinite tower of momentum space soft theorems in \cite{Li:2018gnc}.   We begin by synthesizing various results from the existing literature to present a systematic derivation of the symmetry action on massless hard particles in $(3,1)$ signature.  We demonstrate that this analysis yields the same results as were obtained from $(2,2)$ analysis in \cite{Himwich:2021dau}.  Then, we generalize the $(3,1)$ analysis to massive hard particles. In Section \ref{sec:momentumaction}, we construct charges that generate the symmetries associated to the infinite tower of soft theorems and show that their action on massive particles in momentum space respects the algebra of ${\rm w}_{1+\infty}$. We construct a generating function for the symmetry action at each order in the soft expansion in Section \ref{sec:genfun}, which elucidates the differences between the massless and massive case and facilitates the transformation of the symmetry action to the conformal primary basis. In Section \ref{sec:conformalpb}, we transform our results to the conformal primary basis. Appendix \ref{four-point-section} is dedicated to the derivation of \eqref{non-local-amp} and further explanation of the surrounding discussion in the introduction. Appendix \ref{app_b} provides a derivation of the statement in the introduction that the subset of ${\rm SL}(2, \mathbb{C})$ transformations that preserve a point on the boundary of ${\rm AdS}_3$ preserve no proper subset of points in the bulk of ${\rm AdS}_3$. Further details of the generating function derivation and proof of the ${\rm w}_{1+\infty}$ commutators are presented in Appendix \ref{app:massivegen} and Appendix \ref{app:w52}, respectively.

\section{Review of celestial amplitudes and ${\rm SL}(2, \mathbb{C})$ transformations} \label{sec:review}
    
In this section, we review the construction of celestial amplitudes from scattering amplitudes in momentum space.   In preparation for subsequent sections, we also include the ${\rm SL} (2, \mathbb{C})$ transformation properties of expressions at various intermediate stages of the analysis. 
    
The first step in constructing conformal primary states of massless particles is to parametrize the scattering data, consisting of momenta $p_k$ and polarizations $\ve_k$, by an energy scale $\omega_k$, a point on the 2D celestial plane $(z_k, \bar{z}_k)$, and a sign $\eps_k$ to distinguish outgoing ($\eps_k = +1$) from incoming ($\eps_k = -1$)  particles:
\begin{equation} \label{massless-par}
    \begin{split}
        p^\mu_k = \eps_k\omega_k \hat q^\mu(z_k, \bar{z}_k), \quad \quad \quad 
        \varepsilon_{k+}^\mu = \frac{1}{\sqrt{2}} \frac{1}{z_{0k}} \hat q^\mu(z_0, \bar{z}_k), \quad \quad 
        \varepsilon_{k-}^\mu = \frac{1}{\sqrt{2}} \frac{1}{\bar{z}_{0k}} \hat q^\mu(z_k, \bar{z}_0), 
    \end{split}
\end{equation}  
where
\begin{equation}
    \begin{split} 
        \hat q^\mu(z, \bar{z})
            = \left(1+z \bar{z}, z+\bar{z}, -i(z-\bar{z}), 1-z \bar{z}\right).
    \end{split}
\end{equation}
Here we have introduced an auxiliary point $(z_0, \bar{z}_0)$ so that the polarizations transform covariantly under ${\rm SL}(2, \mathbb{C})$.  This is equivalent to the introduction of a reference vector in the spinor helicity formalism. Explicitly, under the transformation 
\begin{equation} \label{sl2c}
    \begin{split}
        z \to z' =  \frac{az+b}{cz+d}, \quad \quad \quad 
        \bar{z} \to \bar{z}'= \frac{\bar{a}\bar{z}+\bar{b}}{\bar{c}\bar{z}+\bar{d}}, \quad \quad \quad 
        ad-bc = \bar{a}\bar{d}-\bar{b}\bar{c} = 1,
    \end{split}
\end{equation}
we find 
\begin{equation} \label{massless-sl2c}
    \begin{split}
        \hat q^\mu (z, \bar{z}) &\to \hat q^\mu (z', \bar{z}') 
            = \frac{1}{(cz+d)(\bar{c} \bar{z}+ \bar{d})} \Lambda^\mu{}_\nu \hat q^\nu (z,\bar{z}),\\
        \varepsilon_+^\mu(z, \bar{z};z_0)&\to \varepsilon_+^\mu(z', \bar{z}';z_0')
            = \frac{(cz+d)}{(\bar{c} \bar{z}+ \bar{d})} \Lambda^\mu{}_\nu \varepsilon_+^\nu(z, \bar{z};z_0),\\
        \varepsilon_-^\mu(z, \bar{z};\bar{z}_0)&\to \varepsilon_-^\mu(z', \bar{z}';\bar{z}_0')
            = \frac{(\bar{c}\bar{z}+\bar{d})}{(cz+d)} \Lambda^\mu{}_\nu \varepsilon_-^\nu(z, \bar{z};\bar{z}_0),\\
        p^\mu_k(\omega_k, z_k, \bar{z}_k)& \to p^\mu_k(\omega'_k, z'_k, \bar{z}'_k)
             = \Lambda^\mu{}_\nu p^\nu_k(\omega_k, z_k, \bar{z}_k),
    \end{split}
\end{equation}
where $\Lambda$ is the corresponding Lorentz transformation in the vector representation and in the last line we use
\begin{equation}
    \omega'_k = (cz_k+d) (\bar{c} \bar{z}_k + \bar{d}) \omega_k.
\end{equation}
Note that in the limit $z_0, \bar{z}_0 \to \infty$, we recover the familiar expressions for polarization vectors:
\begin{equation}
    \begin{split}
        \varepsilon_{k+}^\mu = \frac{1}{\sqrt{2}} \partial_{z_k} \hat q^\mu (z_k, \bar{z}_k), \quad\quad\quad\quad 
        \varepsilon_{k-}^\mu = \frac{1}{\sqrt{2}} \partial_{\bar{z}_k} \hat q^\mu (z_k, \bar{z}_k).
    \end{split}
\end{equation}  
Polarization tensors for gravitons are built out of the polarization vectors in \eqref{massless-par} by $\varepsilon_{k\pm}^{\mu\nu} = \varepsilon_{k\pm}^{\mu}\varepsilon_{k\pm}^{\nu}$. 
    
After parametrizing the scattering data as described above, conformal primary states of massless particles can be obtained by performing a Mellin transformation with respect to energy scale $\omega_k$:
\begin{equation}
    \int_0^\infty \frac{d \omega_k}{\omega_k}~ \omega_k^{\Delta_k}.
\end{equation}
    
Massive conformal primary states are likewise constructed using a parametrization of massive momenta $p_k$ by a mass $m_k$, a point $\hat p_k(y_k,w_k, \bar{w}_k)$ on the unit hyperboloid $\hat p_k^2= -1$, and a sign $\eps_k$:
\begin{equation} \label{massive-par}
    \begin{split}
        p_k^\mu = \eps_k m_k \hat p_k^\mu, \quad\quad\quad 
        \hat p_k^\mu = \frac{1}{2 y_k} \left(y_k^2 n^\mu + \hat q^\mu (w_k, \bar{w}_k)\right),
    \end{split}
\end{equation}
where
\begin{equation}
    \begin{split}
        n^\mu \equiv \partial_z \partial_{\bar{z}} \hat q^\mu (z, \bar{z}).
    \end{split}
\end{equation} 
${\rm SL} (2, \mathbb{C})$ acts on these parameters like the isometries of ${\rm AdS}_3$ on coordinate charts,
\begin{equation}
    \begin{split}
        w_k \to w'_k & =\frac{(aw_k+b)(\bar{c} \bar{w}_k + \bar{d})+ a \bar{c} y_k^2}{(cw_k+d)(\bar{c} \bar{w}_k + \bar{d})+ c \bar{c} y_k^2},\\
        \bar{w}_k \to \bar{w}'_k & =\frac{(\bar{a}\bar{w}_k+\bar{b})(cw_k+d)+ \bar{a}c y_k^2}{(cw_k+d)(\bar{c} \bar{w}_k + \bar{d})+ c \bar{c} y_k^2},\\
        y_k \to y'_k & =\frac{y_k}{(cw_k+d)(\bar{c} \bar{w}_k + \bar{d})+ c \bar{c} y_k^2},
    \end{split}
\end{equation}
and generates Lorentz transformations on massive momenta:
\begin{equation}
\label{massive-sl2c}
    \begin{split}
        p^\mu_k (y_k, w_k, \bar{w}_k) \to 
                 p^\mu_k (y'_k, w'_k, \bar{w}'_k) = \Lambda^\mu{}_\nu  p^\nu_k (y_k, w_k, \bar{w}_k).
    \end{split}
\end{equation}
    
Conformal primary states of massive scalars are then obtained by integrating the result against an ${\rm AdS}_3$ bulk-to-boundary propagator
\begin{equation}
    \begin{split}
        \mathcal{G}_\Delta(\hat p_k; \hat q_k)
            \equiv \frac{1}{(-\hat p_k \cdot \hat q(z_k, \bar{z}_k))^\Delta}
            = \left(\frac{y_k}{y_k^2 + (w_k-z_k)(\bar{w}_k- \bar{z}_k)}\right)^{\Delta}
    \end{split} 
\end{equation} 
with the ${\rm SL}(2,\mathbb{C})$-invariant measure on the unit three-dimensional hyperboloid 
\begin{equation}
    \begin{split}
        \int [d\hat p_k] = \int_0^\infty \frac{dy_k}{y_k^3} \int d^2 w_k.
    \end{split}
\end{equation} 
This construction readily generalizes to massive spinning particles by simply replacing the scalar bulk-to-boundary propagator with its appropriate spinning counterpart \cite{Law:2020tsg,Iacobacci:2020por,Narayanan:2020amh}.
        
Putting everything together, we obtain the following relation between celestial amplitudes $\langle \mathcal{O} \cdots \rangle$ and momentum-space amplitudes $\mathcal{A}$:
\begin{equation} \label{def-celestialamp}
    \begin{split}
        \langle \mathcal{O}_{\Delta_1}(z_1, \bar{z}_1)\cdots\mathcal{O}_{\Delta_n}(z_n, \bar{z}_n)  \rangle
        = \prod_{\substack{j~\rm massless\\ k~\rm massive}} 
            \int_0^\infty \frac{d \omega_j}{\omega_j} \omega_j^{\Delta_j}
            \int [d\hat p_k] ~\mathcal{G}_{\Delta_k} (\hat p_k; \hat q_k) ~ \mathcal{A}(p_1, \cdots, p_n),
    \end{split}
\end{equation}
where we have suppressed all spin labels on both celestial operators $\mathcal{O}_{\Delta}$ and bulk-to-boundary propagators $\mathcal{G}_{\Delta}$. Under the action of ${\rm SL}(2, \mathbb{C})$ \eqref{sl2c}, celestial amplitudes transform according to  
\begin{equation} \label{sl2ctrans}
    \begin{split}
        \langle \mathcal{O}_{\Delta_1 }  (z'_1, \bar{z}'_1) \cdots\mathcal{O}_{\Delta_n } (z'_n, \bar{z}'_n)  \rangle 
        = \left[\prod_{k = 1}^n (cz_k+d)^{2h_k}(\bar{c} \bar{z}_k+\bar{d})^{2\bar{h}_k} \right]
        \langle \mathcal{O}_{\Delta_1 }  (z_1, \bar{z}_1) \cdots\mathcal{O}_{\Delta_n } (z_n, \bar{z}_n)  \rangle,
    \end{split}
\end{equation}
where $(h_k, \bar{h}_k)$ are the left and right conformal weights, related to $\Delta_k$ and $s_k$ by
\begin{equation}
    \begin{split}
        (h_k, \bar{h}_k) = \left(\frac{\Delta_k+s_k}{2}, \frac{\Delta_k-s_k}{2}\right).
    \end{split}
\end{equation}
        
The analysis in the following sections will focus on amplitudes for scalar particles and positive helicity (conformally) soft gravitons.  We introduce $G_{\Delta}$ to denote a positive helicity graviton of conformal weight $\Delta$ and $\Phi_\Delta$ ($\phi_\Delta$) to denote massive (massless) conformal primary scalars of weight $\Delta$. In expressions where mass is irrelevant, $\Phi_\Delta$ and $\phi_\Delta$ are collectively represented by $\mathcal{O}_\Delta$. We expect that all non-trivial aspects of the analysis appear in the coupling of (conformally) soft gravitons to scalar particles and thus the generalization of the following analysis to spinning particles will be straightforward. 
     
\section{From soft theorems to ${\rm w}_{1+\infty}$} \label{sec:softw}
 
In \cite{Li:2018gnc}, the authors use gauge invariance to show that an amplitude for graviton emission admits universal behavior to all orders in the low-energy expansion
\begin{equation}
    \begin{split}
        \mathcal{A} (\omega \hat q; p_1, \cdots ,p_n)
             = \frac{\kappa}{2}\sum_{\ell = -1}^\infty \omega^\ell \mathcal{A}^{(\ell)} ( \hat q; p_1, \cdots ,p_n),
    \end{split}
\end{equation}
where $\kappa^2 = 32 \pi G$. The leading and subleading terms in the expansion take the previously-discovered \cite{Weinberg:1965nx,Cachazo:2014fwa} form 
\begin{equation} \label{leading}
    \begin{split}
        \mathcal{A}^{(-1)} ( \hat q; p_1, \cdots ,p_n) 
        = \sum_{k = 1}^n S''^{(-1)}_k \mathcal{A} ( p_1, \cdots ,p_n)  ,
        \quad \quad \quad S''^{(-1)}_k = \frac{\ve_{\mu\nu}  p_k^\mu p_k^\nu}{ \hat q \cdot   p_k},
    \end{split}
\end{equation} 
and 
\begin{equation}\label{subleading}
    \begin{split}
        \mathcal{A}^{(0)} ( \hat q; p_1, \cdots ,p_n)= \sum_{k = 1}^n S''^{(0)}_k
        \mathcal{A} ( p_1, \cdots ,p_n),\quad \quad \quad S''^{(0)}_k  = \frac{\ve_{\mu\nu} p_k^\mu (i\hat q_\sigma \mathcal{J}^{\nu \sigma}_k) }{ \hat q \cdot  p_k},
    \end{split}
\end{equation} 
respectively.  Here the soft graviton has momentum $\omega \hat q$ and polarization $\varepsilon_{\mu\nu}$ and $p_k$, $\mathcal{J}_k$ are the momentum and total angular momentum of the $k$th particle. Both orbital $\mathcal{L}_k$ and intrinsic $\mathcal{S}_k$ angular momentum contribute to $\mathcal{J}_k$:
\begin{equation}
    \begin{split}
        \mathcal{J}_k  = \mathcal{L}_k+\mathcal{S}_k, \quad \quad \quad 
        \mathcal{L}_k{}_{\mu\nu} = -i \left(p_{k\mu} \frac{\partial}{\partial p_k^\nu}-p_{k\nu} \frac{\partial}{\partial p_k^\mu}\right).
    \end{split}
\end{equation} 
Beyond subleading order, the soft theorems were found to take the form \cite{Hamada:2018vrw,Li:2018gnc}
\begin{equation}\label{subnleading}
    \begin{split}
        \mathcal{A}^{(\ell > 0)}(\hat q; p_1,\cdots,p_n) & = \sum_{k = 1}^n S''^{(\ell)}_k
        \mathcal{A}( p_1, \cdots ,p_n) +\ve_{\mu\nu} \mathcal{B}^{\mu\nu}_{\ell} (\hat q; p_1, \cdots, p_n), 
    \end{split}
\end{equation}  
where 
\begin{equation} \label{eq:LLZsoft}
    \begin{split} 
         S''^{(\ell)}_k  =\frac{1}{(\ell+1)!} \frac{\ve_{\mu\nu} (i\hat q_\rho \mathcal{J}^{\mu \rho}_k)(i\hat q_\sigma \mathcal{J}^{\nu \sigma}_k) }{\hat q \cdot p_k}
        \left(\hat q  \cdot \frac{\p}{\p p_k}\right)^{\ell-1},  \quad \quad \quad  \ell \in \mathbb{Z}_{\geq 1}.
    \end{split}
\end{equation}
Here $\mathcal{B}^{\mu\nu}_{\ell}$ is a non-universal contribution, which is generally non-vanishing.  The non-universal corrections at subsubleading order ($\ell = 1$) are controlled at tree level by a finite number of curvature couplings (including $RF^2$ and $R^2 \Phi$) \cite{Elvang:2016qvq}. Analogous structure beyond subsubleading order is not known nor necessarily expected. 
 
Equivalent statements for celestial amplitudes are obtained by performing the appropriate change of basis, reviewed in the previous section.  Using the mathematical identity\footnote{\eqref{omega-identity} converges provided that the amplitude falls off at high energies as $\mathcal{A}(\omega) \sim \omega^{-m}$ where $m+\ell > \eps$.}
\begin{equation} \label{omega-identity}
    \begin{split}
        \lim_{\omega \to 0} \p_{\omega}^{\ell +1} \left(\omega \mathcal{A}(\omega)\right) =
        (\ell+1)! \lim_{\epsilon \to 0} \epsilon \int_0^\infty \frac{d\omega}{\omega}~\omega^{-\ell+\epsilon} \mathcal{A}(\omega), 
    \end{split} 
\end{equation} 
we recall that the soft limit takes the following form in the conformal primary basis \cite{Cheung:2016iub,Fan:2019emx,Pate:2019mfs,Nandan:2019jas, Adamo:2019ipt,Puhm:2019zbl,Guevara:2019ypd}
\begin{equation} \label{softlim-cpb}
    \begin{split}
        \langle H_{-\ell}(z,\bar{z}) \mathcal{O}_{\Delta_1}(z_1, \bar{z}_1) \cdots&\mathcal{O}_{\Delta_n} (z_n, \bar{z}_n) \rangle 
        \\&=\prod_{\substack{j~\rm massless\\ k~\rm massive}} \int_0^\infty \frac{d \omega_j}{\omega_j} \omega_j^{\Delta_j}
        \int [d\hat p_k] ~\mathcal{G}_{\Delta_k} (\hat p_k; \hat q_k) \mathcal{A}^{(\ell)} (\hat q; p_1, \cdots ,p_n),
    \end{split}
\end{equation}
where
\begin{equation}
    \begin{split}
        H_{-\ell}(z, \bar{z}) \equiv \lim_{\epsilon \to 0} \epsilon  G_{-\ell+\epsilon}(z,\bar{z}),  
    \end{split}
\end{equation}
is a conformally soft graviton that transforms under ${\rm SL}(2, \mathbb{C})$ with  conformal weight
\begin{equation} \label{current-weight}
    \begin{split}
        (h, \bar{h}) = \left(\frac{-\ell+2}{2},\frac{-\ell-2}{2} \right).
    \end{split}
\end{equation}  
Notice that the soft graviton in \eqref{softlim-cpb} is already in a state with definite conformal weight. This equivalence between momentum-space soft limits and emissions of particles of definite conformal weight has several non-trivial consequences.  
  
First, it clarifies the conditions under which non-universal corrections $\mathcal{B}^{\mu\nu}_\ell$ can appear. To see this, treat $z$ and $\bar{z}$ as independent complex variables and consider the following celestial amplitude involving only massless celestial operators  $\phi_{\Delta}$:
\begin{equation} 
\label{contour-deformation-453}
    \begin{split}
        \langle H_{-\ell}&(z,\bar{z}) \phi_{\Delta_1}  (z_1, \bar{z}_1) \cdots  \rangle\\
        & = \oint_{z} \frac{dw}{2\pi i}\frac{1}{w-z} \langle H_{-\ell}(w,\bar{z}) \phi_{\Delta_1}  (z_1, \bar{z}_1) \cdots \rangle\\
        & = - \sum_{j = 1}^m \oint_{z_j} \frac{dw}{2 \pi i}\frac{1}{w-z} \langle H_{-\ell}(w,\bar{z}) \phi_{\Delta_1 }  (z_1, \bar{z}_1) \cdots \rangle + \oint_{\infty} \frac{dw}{2 \pi i}\frac{1}{w-z} \langle H_{-\ell}(w,\bar{z}) \phi_{\Delta_1}(z_1, \bar{z}_1)\cdots  \rangle.
    \end{split}
\end{equation}
To reach the last line, we have exploited the fact that in the soft limit, the delta-function for momentum conservation no longer constrains the position $w$ of the soft graviton and the amplitude is thus meromorphic in $w$.  This allows us to deform the contour to infinity at the price of including contributions from  singularities at other points, denoted collectively by $z_j$.  As mentioned in the introduction and detailed in Appendix \ref{four-point-section}, the assumption of meromorphicity is only strictly justified for massless scattering.
        
Next, we exploit ${\rm SL}(2,\mathbb{C})$ symmetry to study the contribution at infinity. Using \eqref{sl2ctrans}, we find that under a general conformal transformation \eqref{sl2c}, the celestial amplitude transforms as 
\begin{equation} 
\label{trans1}
    \begin{split}
        \langle H_{-\ell}(z',\bar{z}')\phi_{\Delta_1}(z_1',\bar{z}_1')\cdots\phi_{\Delta_n}(z_n',\bar{z}_n') \rangle
        &= (cz+d)^{-\ell+2}(\bar{c} \bar{z}+\bar{d})^{-\ell-2} \prod_{k=1}^n [(c z_k+d)(\bar{c}\bar{z}_k+\bar{d})]^{\Delta_k} 
        \\& \quad\quad\quad\quad\quad\times
        \langle H_{-\ell}(z,\bar{z})\phi_{\Delta_1}(z_1, \bar{z}_1) \cdots\phi_{\Delta_n}(z_n,\bar{z}_n)\rangle.
    \end{split}
\end{equation}
Notice that the limit $z \to \infty$ sends $z' \to a/c$.  Since $a$ and $c$ are generic, we expect the right-hand side of \eqref{trans1} to be finite when evaluated in this limit at $z' = a/c$.  Hence, we deduce that 
\begin{equation} 
\label{conformal-fall-off}
    \begin{split}
        \lim_{z \to \infty} \langle H_{-\ell}(z,\bar{z}) \phi_{\Delta_1}(z_1,\bar{z}_1)\cdots\phi_{\Delta_n}(z_n, \bar{z}_n) \rangle \sim \frac{1}{z^{-\ell+2}}.
    \end{split}
\end{equation}

The leading, subleading, and subsubleading soft limits have $\ell = -1, 0, 1$, respectively. In these cases the contribution from the contour around infinity in \eqref{contour-deformation-453} thus vanishes and the conformally soft limit is entirely determined by the singularities in $z$.  In celestial amplitudes for massless scattering, the singularities in $z$ are entirely dictated by the collinear singularities in momentum space amplitudes or equivalently the singular terms in the celestial operator product expansion (OPE). Indeed, one can verify that these precisely reproduce the soft factors in \eqref{leading} and \eqref{subleading}, the leading and subleading case respectively.  At subsubleading order, the celestial OPE can receive corrections from a finite number of curvature couplings (including $RF^2$ and $R^2 \Phi$), but still precisely reproduces the corrections to the subsubleading soft graviton theorem that arise from these couplings. Thus, in purely massless scattering, the form of the soft theorems up to subsubleading order is tightly controlled by ${\rm SL}(2, \mathbb{C})$ symmetry.  In particular,  ${\rm SL}(2, \mathbb{C})$ symmetry implies factorization, namely that the amplitude for soft emission can be written as a soft factor (or operator) acting on the amplitude for a hard scattering process. As we saw in the introduction, the singularity structure is more subtle when massive particles are present. Nevertheless, conformal symmetry ensures a sufficiently fast fall-off in the limit $z \to \infty$ so that the contribution from $z = \infty$ to a contour integral of $H_{-\ell}$ vanishes for $\ell = -1, 0, 1$.  Therefore, up to and including subsubleading order, the conformally soft limit is still fully determined by the singularity structure in $z$.\footnote{Taking a contour integral in $z$ of \eqref{contour-deformation-453} that surrounds all of the hard insertions, a similar argument implies that the leading and subleading soft graviton insertions with $\ell = -1,0$ correspond to globally conserved charges (the familiar global translations and Lorentz transformations). The analog of this statement for global $U(1)$ charge conservation is presented in Appendix \ref{four-point-section} in the derivation of \eqref{eval-2}. However, at subsubleading order and beyond ($\ell \geq 1$) the contour integrals include terms from infinity and therefore the soft insertions at these orders do not correspond to globally conserved charges.}

On the other hand, when $\ell \geq 2$ (equivalently beyond subsubleading order), the contribution from the contour around infinity in \eqref{contour-deformation-453} need not vanish.  This means that the conformally soft limit is not entirely determined by the singularities in $z$, and thus for massless particles not entirely determined by the collinear singularities of momentum-space amplitudes, nor equivalently by the celestial OPE. This result is fully consistent with the non-universal corrections in \eqref{subnleading} that were found in \cite{Li:2018gnc} to appear beyond subsubleading order and need not admit the factorizing form described above. 

This result further demonstrates that statements about universal soft behavior beyond subsubleading order require more care.  In particular, the presence of a non-universal contribution introduces an ambiguity in defining the universal contribution because terms can in principle be freely moved between the two. The authors in \cite{Li:2018gnc} address this ambiguity by defining the universal contribution as the image of a particular projection operator.

In this work, we demonstrate that the ambiguity can be resolved such that the infinite tower of soft graviton theorems organizes according to a ${\rm w}_{1+\infty}$ symmetry.  The manifest ${\rm SL}(2, \mathbb{C})$-covariance of scattering amplitudes in a conformal primary basis was instrumental in the discovery of the original formulation of ${\rm w}_{1+\infty}$ in celestial holography \cite{Guevara:2021abz,Himwich:2021dau,Strominger:2021lvk}. We therefore seek expressions for the soft factor that render the partition into universal and non-universal pieces invariant under the action of ${\rm SL}(2, \mathbb{C})$. Since the conformally soft graviton transforms like a primary of weight $(h, \bar{h}) = \left(\frac{-\ell+2}{2},\frac{-\ell-2}{2} \right)$, a conformally-invariant partition thus requires the associated soft factor also to transform as a primary with the same conformal weight. Upon identifying expressions that transform like ${\rm SL}(2, \mathbb{C})$ primaries, we further demonstrate that these generate a ${\rm w}_{1+\infty}$ symmetry action on hard massless and massive particles.  Our derivation of a ${\rm w}_{1+\infty}$ symmetry action on massless hard particles from momentum-space soft theorems merely synthesizes a variety of existing results from the literature into a systematic procedure.  On the other hand, the generalization to massive particles reveals new expressions for universal soft behavior that to our knowledge have not yet appeared in the literature. 

\subsection{Massless particles} \label{sec:massless}
       
The soft factors appearing in \eqref{eq:LLZsoft} do not transform like primaries. Specifically, the factors of $\hat q \cdot \partial_{p_k}$ are problematic. However, a natural proposal is to complete these factors into angular momentum generators:
\begin{equation} \label{angmom-completion}
    \begin{split}
        \hat q^\mu \frac{\partial}{\partial p_k^\mu} \to 
            \frac{1}{\varepsilon_+ \cdot p_k} \left( (\varepsilon_+ \cdot p_k)~ \hat q^\mu \frac{\partial}{\partial p_k^\mu}-  (\hat q   \cdot p_k)~ \varepsilon_+^\mu \frac{\partial}{\partial p_k^\mu}\right)
        = \frac{1}{\varepsilon_+ \cdot p_k}~ F_{+}^{\mu \nu} \mathcal{L}_k{}_{\mu \nu},
    \end{split}
\end{equation}
where $F_\pm$ is the field strength 
\begin{equation}
    \begin{split} 
        F_\pm^{\mu \nu}  = i  \varepsilon_\pm^{[\mu}  \hat q^{\nu]}.
    \end{split}
\end{equation}
Notice that division by $\varepsilon_+ \cdot p_k$ renders the expression gauge-dependent (i.e.~non-invariant under $\varepsilon_+ \to \varepsilon_+ + \hat q$).  We will show that this gauge dependence ultimately drops out from the symmetry action. 

Under ${\rm SL}(2, \mathbb{C})$ transformations \eqref{sl2c} in which the reference point $z_0$ is also transformed, the modified soft factors  
\begin{equation}\label{softfactor1}
    \begin{split}
        S'^{(\ell)}_k( z,\bar{z}) =
            \frac{\varepsilon_+{}_{\mu\nu} p^\mu_k p^\nu_k}{\hat q \cdot p_k} \frac{1}{(\ell+1)!} \left(\frac{F_+ \cdot \mathcal{J}_k}{\varepsilon_+ \cdot p_k}\right)^{\ell+1} 
    \end{split}
\end{equation}
transform like primaries, which we demonstrate shortly.  In \eqref{softfactor1}, 
\begin{equation} \label{softpar}
     \hat q^\mu = \hat q^\mu (z, \bar{z}), \quad \quad \quad 
     \varepsilon^\mu_+ = \varepsilon^\mu_+ (z, \bar{z}; z_0), \quad \quad \quad 
    F_+ \cdot \mathcal{J}_k = F_{+}^{\mu\nu }\mathcal{J}_k{}_{\mu\nu },
\end{equation}
and we have further replaced the orbital angular momentum $\mathcal{L}_k$ with the total angular momentum $\mathcal{J}_k$.  These are equivalent for scalars, which are the focus of the present analysis, but we expect that \eqref{softfactor1} is also the correct generalization for spinning particles.  In fact, \eqref{softfactor1} has appeared before in the literature \cite{Bautista:2019tdr,Adamo:2021lrv}. The authors of \cite{Bautista:2019tdr} observe that this expression implies that the full soft factor expansion resums into an exponential:
\begin{equation}
     \sum_{\ell=-1}^\infty \omega^{\ell}   S'^{(\ell)}_k( z,\bar{z}) 
    = \frac{\varepsilon_+{}_{\mu\nu} p^\mu_k p^\nu_k}{ \omega \hat q \cdot p_k} \exp \left( \omega \frac{F_+ \cdot \mathcal{J}_k}{   \varepsilon_+ \cdot p_k}\right). 
\end{equation}

An additional advantage of \eqref{softfactor1} is that it subsumes the leading and subleading cases and applies to every order in the soft expansion ($\ell \geq -1$). Therefore, we can verify the transformation of the tower of soft factors under ${\rm SL}(2, \mathbb{C})$ all in one go.  In particular, using \eqref{massless-sl2c} and \eqref{massive-sl2c}, one finds 
\begin{equation}
    \begin{split}
        \frac{\varepsilon_+{}_{\mu\nu} p^\mu_k p^\nu_k}{\hat q \cdot p_k}& \to \frac{(cz+d)^3}{(\bar{c}\bar{z}+\bar{d})}
        ~\frac{\varepsilon_+{}_{\mu\nu} p^\mu_k p^\nu_k}{\hat q \cdot p_k}, \\
       \varepsilon_+ \cdot p_k &\to \frac{(cz+d)}{(\bar{c}\bar{z}+\bar{d})}~\varepsilon_+ \cdot p_k, \\
        F_+ \cdot \mathcal{L}_k & \to \frac{1}{(\bar{c}\bar{z}+\bar{d})^2}~F_+ \cdot \mathcal{L}_k.
    \end{split}
\end{equation}
Putting these together, the total transformation of the proposed soft factor  
\begin{equation}
    \begin{split} \label{masslesstsf}
         S'^{(\ell)}_k (z, \bar{z})  \to  
             (cz+d)^{-\ell+2}(\bar{c}\bar{z}+\bar{d})^{-\ell-2}S'^{(\ell)}_k  (z, \bar{z}),
    \end{split}
\end{equation}
precisely matches the transformation of a weight $(\frac{-\ell+2}{2}, \frac{-\ell-2}{2})$ primary.

Notice that \eqref{masslesstsf} follows from transforming the reference point $z_0$ in addition to the points that specify the momenta of the scattering particles.  Since the reference point is arbitrary and the full amplitude is independent of its choice, the ``non-universal" piece beyond subsubleading order must also depend on the reference point.  At leading and subleading order, momentum and angular momentum conservation can be used to show that the choice of reference point drops out in the sum over $k$ in \eqref{leading} and \eqref{subleading}, respectively, while at subsubleading order $S'^{(\ell=1)}_k$  is simply independent of the reference point.  

Strictly speaking, the non-trivial dependence on the reference vector for $\ell > 1$ indicates that the partition is not conformally  invariant. Explicitly, when the reference vector is not transformed, it is clear that the partition is not preserved.  This is equivalent to the previous observation that the factors of $\varepsilon_+ \cdot p_k$ appearing in the denominator of $S'^{(\ell)}_k (z, \bar{z})$ when $\ell >1$ render the soft factor non-invariant under gauge transformations $\varepsilon_+ \to \varepsilon_+ + \hat q$.  Nevertheless, we now show that our analysis of massless hard particles tolerates this subtle non-invariance because it can be removed in a conformally covariant manner.

To do so, we exploit the property of conformally soft positive-helicity gravitons that they admit right-moving descendants that are also primaries (i.e.~primary-descendants) \cite{Banerjee:2019aoy,Banerjee:2019tam,Guevara:2021abz,Pasterski:2021dqe,Pasterski:2021fjn}. By virtue of respecting ${\rm SL} (2, \mathbb{C})$ transformations identical to those of conformally soft gravitons, $S'^{(\ell)}_k (z, \bar{z})$ is likewise guaranteed to admit a primary descendant.   That is, under ${\rm SL}(2, \mathbb{C})$, the soft factor descendant $\partial^{\ell+3}_{\bar{z}}S'^{(\ell)}_k (z, \bar{z})$ transforms like a primary of weight $(\frac{-\ell+2}{2},  \frac{\ell+4}{2})$: 
\begin{equation}
    \begin{split}
        \partial^{\ell+3}_{\bar{z}}S'^{(\ell)}_k (z, \bar{z})
        \to   (cz+d)^{-\ell+2}(\bar{c}\bar{z}+\bar{d})^{\ell+4} \partial^{\ell+3}_{\bar{z}}S'^{(\ell)}_k  (z, \bar{z}),
    \end{split}
\end{equation}
which follows from the following identity for any function $f(\bar{z})$:
\begin{equation} \label{primdesc-id}
    \begin{split}
        \left[(\bar{c}\bar{z}+ \bar{d})^2 \partial_{\bar{z}}\right]^{\ell+3}
            \left[\frac{f(\bz)}{(\bar{c}\bar{z}+ \bar{d})^{\ell+2}} \right]
            =(\bar{c}\bar{z}+ \bar{d})^{\ell+4}\partial_{\bar{z}}^{\ell+3} f(\bar{z}).
    \end{split}
\end{equation}

Rewriting \eqref{softfactor1} using \eqref{softpar} and the parametrization for massless particles \eqref{massless-par}, we obtain the following explicit expression: 
\begin{equation} 
    \begin{split}
        S'^{(\ell)}_k  (z, \bar{z})& =- \frac{  \epsilon_k\omega_k}{(\ell+1)!}
           \frac{ \bar{z}- \bar{z}_k}{z-z_k} \left(\frac{z-z_0}{z_k-z_0}\right)^{\ell-1} \left(  \frac{1}{ \epsilon_k\omega_k} \left[\omega_k \partial_{\omega_k}-(\bar{z}_k- \bar{z}) \partial_{\bar{z}_k}\right]\right)^{\ell+1}\\
         & =   \frac{(-1)^\ell}{(\ell+1)!}\frac{  \bar{z}-\bar{z}_k }{ z-z_k } \left( \frac{z-z_0}{z_k-z_0}\right)^{ \ell-1}
		\left( \sum_{m = 0}^{\ell+1} { \ell+1 \choose m}   \frac{\Gamma(-\omega_k\partial_{\omega_k} +1)}{\Gamma(-\omega_k\partial_{\omega_k} -\ell+m)} (\bar{z}_k -\bar{z})^m\partial_{\bar{z}_k}^m  \right) \left(\epsilon_k\omega_k\right)^{-\ell} .
    \end{split} 
\end{equation}
To arrive at this expression, it is helpful to notice that
\begin{equation} \label{eq:FLmassless}
	\begin{split}
		F_+ \cdot  \mathcal{L}_k 
			& = -\sqrt{2} \left[(\bar{z}_k-\bar{z} )^2  \partial_{\bar{z}_k} - (\bar{z}_k- \bar{z} ) \omega_k \partial_{\omega_k}  \right] ,\\
		F_- \cdot  \mathcal{L}_k 
			& = -\sqrt{2} \left[(z_k-z )^2\partial_{z_k} - (z_k- z ) \omega_k \partial_{\omega_k}  \right] .
	\end{split}
\end{equation}
Then, expression for the associated primary descendant takes the explicit form 
\begin{equation} \label{masslesspd}
    \begin{split}
      \partial_{\bar{z}}^{\ell+3}  S'^{(\ell)}_k  (z, \bar{z}) =    \sum_{m = 0}^{\ell+1} \frac{(-1)^{\ell+m}  (m+1)  }{(\ell+1-m)! }  \frac{\Gamma(-\omega_k\partial_{\omega_k} +1)}{\Gamma(-\omega_k\partial_{\omega_k} -\ell+m)}  2 \pi \partial_{\bar{z}}^{\ell+1-m} \delta^{(2)}(z-z_k) ~\partial_{\bar{z}_k}^m   \left(\epsilon_k\omega_k\right)^{-\ell} .
    \end{split}
\end{equation}
Importantly, this descendant no longer depends on the reference point $z_0$.  This implies that -- so long as the modes of the primary descendant can be shown to generate the action ${\rm w}_{1+\infty}$ -- we need not identify a conformally invariant partition at the level of the soft factor.  In other words, we have obtained a conformally invariant partition at the level of the primary descendant of the soft factor.  Of course, it may be possible to modify the massless soft factor from \cite{Li:2018gnc} to be strictly conformally covariant or equivalently independent of choice of reference point $z_0$.  However, as we will now demonstrate, it is unnecessary to find such an expression for the sole purpose of establishing a ${\rm w}_{1+\infty}$ symmetry action.

A central lesson from the equivalence between soft theorems and Ward identities for infinite-dimensional symmetries is the interpretation of the universal soft factor as the generator of an infinitesimal symmetry transformation on hard, single-particle states:
\begin{equation}
        \begin{split}
            S^{(\ell)}_k (z, \bar{z})| p_k \rangle \sim \delta^{(\ell)}_{(z , \bar{z})} |p_k\rangle.
        \end{split}
\end{equation}
The underlying symmetry algebra can be extracted by decomposing these local symmetry transformations in a basis of modes and calculating the commutation relations between the modes.  A proper labelling of these modes was essential to the previous discovery \cite{Strominger:2021lvk} of an underlying ${\rm w}_{1+\infty}$ symmetry.

To the same end, we now show that modes of the primary descendant correctly reproduce the action of ${\rm w}_{1+\infty}$ on massless particles found in \cite{Himwich:2021dau}.  Explicitly, consider
\begin{equation} \label{def-wsym}
    \begin{split}
        \delta^p_m |p_k \rangle \equiv  -\frac{1}{2}\int \frac{d^2z}{2 \pi }~ \bar{z}^{p+m-1} \partial_{\bar{z}}^{2p-1}
                    S'^{(2p-4)}_k (z, \bar{z})| p_k \rangle,
    \end{split}
\end{equation}
where
\begin{equation} \label{mode-range}
    \begin{split}
        p = 1, \frac{3}{2}, 2, \frac{5}{2}, \cdots, \quad \quad \quad 1-p \leq m \leq p-1.
    \end{split}
\end{equation}
Here, in anticipation of the results in \cite{Strominger:2021lvk} and \cite{Himwich:2021dau}, we have labelled the transformations by the right-moving conformal weight $p = \bar{h} = \frac{\ell+4}{2}$ and right-moving ${\rm SL} (2, \mathbb{C})$ mode $m$ of the primary descendant.

To compare directly with the results in \cite{Himwich:2021dau}, we need to work in a conformal primary basis. Using \eqref{masslesspd}, we immediately find the action on outgoing particles ($\epsilon_k = 1$) to be
\begin{equation} \label{eq:oldaction}
    \begin{split}
        \delta^q_n \phi_{\Delta_k}(z_k, \bar{z}_k) 
            & =   \int_0^\infty \frac{d \omega_k }{\omega_k} ~ \omega_k^{\Delta_k} ~\delta^q_n |p_k \rangle \\
            &=  \frac{1}{2}  \sum_{m = 0}^{2q-3}{q+n-1\choose m}
	        \frac{(2q-2-m)\Gamma (\Delta_k+1)}{\Gamma (\Delta_k+1-m)}\bar{z}_k^{q+n-1-m} 
	        \partial_{\bar{z}_k}^{2q-3-m}  \phi_{\Delta_k-2q+4}(z_k, \bar{z}_k), 
    \end{split}
\end{equation}
which precisely matches the action of ${\rm w}_{1+\infty}$ found in \cite{Himwich:2021dau}.  Then, by directly importing the results in \cite{Himwich:2021dau}, we reach the conclusion that the symmetry transformations defined in \eqref{def-wsym} respect the algebra  
\begin{equation} \label{w-algebra}
    \left[ \delta^p_m,  \delta^q_n\right]
        = \left[m (q-1)-n(p-1)\right]  \delta^{p+q-2}_{m+n}.
\end{equation}

\subsection{Massive particles} \label{sec:massive}
        
The massless analysis in the previous subsection can only be directly extended to massive particles provided that analogous simplifications arise.  In particular, the observation that the primary descendant was independent of the reference point $z_0$, and thereby constituted a conformally  covariant expression for universal soft behavior, was a critical component of the argument. 

However, when $p_k$ is timelike, the primary descendant of \eqref{softfactor1} is independent of $z_0$ only in the leading few cases. Specifically, when $ \ell = -1, 0, 1$, we find
\begin{equation} \label{pd-1}
    \begin{split}
        \partial_{\bar{z}}^{\ell+3} S'^{(\ell)}_k (z, \bar{z})
            &= \mathcal{N}_\ell \frac{p_k^4}{(\hat q \cdot  p_k)^{\ell+4}} \left(F_{-}\cdot  \mathcal{J}_k \right)^{\ell+1},  
    \end{split}
\end{equation}
where $\mathcal{N}_\ell$ is just a normalization factor
\begin{equation}
    \mathcal{N}_\ell = (-\sqrt{2})^{\ell-1}(\ell+3)(\ell+2).
\end{equation}
To verify $z_0$-independence of \eqref{pd-1}, note that with the parametrization of massive momenta \eqref{massive-par}, the angular momentum generators are $z_0$-independent and take the form 
\begin{equation} \label{massive-angmom}
    \begin{split}
	F_{+}\cdot \mathcal{L}_k 
		&=-  \sqrt{2}  \left[ (\bar{w}_k-\bar{z} )^2  \partial_{\bar{w}_k} + (\bar{w}_k-\bar{z}) y_k \partial_{y_k}  - y_k^2 \partial_{w_k} \right] ,\\
	F_{-} \cdot  \mathcal{L}_k 
		&= -\sqrt{2}  \left[ (w_k-z )^2  \partial_{w_k} + (w_k-z) y_k \partial_{y_k} - y_k^2 \partial_{\bar{w}_k} \right].
    \end{split}
\end{equation}
It then directly follows that \eqref{pd-1} is independent of $z_0$.
	
On the other hand, when $\ell>1$, the primary descendant of \eqref{softfactor1} depends on $z_0$ and thus is not conformally covariant.  To see this explicitly, consider  $\ell=2$.  In this case, using 
\begin{equation} \label{completeness1}
    1 = \frac{p_k^2}{p_k^2} = \frac{(\partial_z \hat q \cdot p_k)(\partial_{\bar{z}} \hat q \cdot p_k)- (\hat q \cdot p_k)(n \cdot p_k) }{p_k^2}
\end{equation}
we can split $S'^{(2)}_k (z, \bar{z})$ into a sum of individually conformally covariant terms, one of which depends on the reference point and one which does not:
\begin{equation} \label{sub3}
    \begin{split}
	S'^{(2)}_k (z, \bar{z})
	    &= \frac{\mathcal{N}_2}{6!} \left[
	            \frac{1}{p_k^2}( \hat q \cdot p_k)^6 \p_{\bar{z}} \left( \frac{( F_+\cdot \mathcal{L}_k )^3}{( \hat q  \cdot p_k)^6}\right)-   \frac{1}{p_k^2}( \hat q_0 \cdot p_k)^6 \p_{\bar{z}} \left( \frac{( F_+\cdot \mathcal{L}_k )^3}{( \hat q_0 \cdot p_k)^6}\right) \right], 
    \end{split}
\end{equation}
where 
\begin{equation}
    \begin{split}
	\hat q_0 \equiv \hat q(z_0, \bar{z}).
    \end{split}
\end{equation} 
The conformal covariance of each individual term can be verified using the following ${\rm SL}(2, \mathbb{C})$ transformations of the constituent pieces:
\begin{equation} \label{sl2c-identities1}
    \begin{split}
        \hat q \cdot   p_k   \to \frac{\hat q \cdot   p_k}{(cz+d)(\bar{c}\bar{z}+ \bar{d})},  \quad \quad \quad 
        F_+ \cdot \mathcal{J}_k    \to \frac{F_+ \cdot \mathcal{J}_k }{(\bar{c} \bar{z}+\bar{d})^2}.
    \end{split}
\end{equation}
Since each term individually transforms like a primary, each term individually admits a primary descendant.  Explicitly, 
\begin{equation}
    \begin{split}
	\partial^{5}_{\bar{z}} \left[ \frac{\mathcal{N}_2}{6!}\frac{1}{p_k^2}( \hat q \cdot p_k)^6 \p_{\bar{z}} \left( \frac{( F_+\cdot  \mathcal{L}_k )^3}{( \hat q  \cdot p_k)^6}\right) \right]
	        = \mathcal{N}_2 \frac{p_k^4}{(\hat q \cdot   p_k)^{6}} \left( F_- \cdot \mathcal{L}_k \right)^{3}, 
    \end{split}
\end{equation}
which is the natural generalization of \eqref{pd-1} to $\ell=2$.  Similarly, the descendant relation for the second term in \eqref{sub3} takes the same form only with $\hat q$ replaced by $\hat q_0$. 
	
More generally, at any order $(\ell \geq 1)$ the soft expression from the previous subsection can be written as a sum over individually conformally covariant terms, only one of which is independent of the reference point:
\begin{equation} \label{massive-exp}
    \begin{split}
	S'^{(\ell)}_k(z, \bar{z})
		&=  \frac{ \mathcal{N}_\ell }{(2\ell+2)!} f^{2\ell+2} \frac{1}{p_k^{2\ell-2}}\p_\bz^{\ell-1}
		\left[\frac{( F_+\cdot  \mathcal{L}_k )^{\ell+1}  }{f^{\ell+4} }\right] \\ & \quad 
		- \frac{ \mathcal{N}_\ell }{(\ell+3)!(\ell-2)! }\frac{1}{p_k^{2\ell-2}}\sum_{j =0}^{\ell-2} {\ell-2 \choose j}\frac{(-1)^j }{j+\ell+4} f^{\ell-2-j} f_0^{\ell+4+j}
		\partial_{\bz}^{\ell-1} \left[\frac{f^j (  F_+\cdot  \mathcal{L}_k )^{\ell+1}  }{f_0^{\ell+4+j}}\right].  
    \end{split}
\end{equation}
Here we have introduced the shorthand 
\begin{equation}
    \begin{split}
	f \equiv \hat q \cdot p_k, 
	          \quad  \quad \quad 
	f_0 \equiv \hat q_0 \cdot p_k .
    \end{split}
\end{equation} 
As in \eqref{sub3}, the explicit division by $p_k^2$ indicates that the massless limit requires some care.  Specifically, it requires that we assemble various contractions of $p_k$ as in the numerator in \eqref{completeness1} and evaluate $p_k^2/p_k^2 = 1$ before taking $p_k^2 \to 0$.  Also as in \eqref{sub3}, the conformal covariance of each term in \eqref{massive-exp} can be verified using \eqref{sl2c-identities1} and \eqref{primdesc-id}.

Since the first term in \eqref{massive-exp} is conformally covariant by itself and independent of the reference point, it is the natural proposal for the universal soft behavior at sub$^{\ell+1}$leading order for massive particles:
\begin{equation} \label{massive-sf}
    \begin{split}
	S^{(\ell)}_k(z, \bar{z})
		&= \frac{\mathcal{N}_\ell}{(2\ell+2)!}  \frac{(\hat q \cdot p_k)^{2\ell+2} }{p_k^{2\ell-2}}  \partial_{\bar{z}}^{\ell-1} \left[\frac{( F_+\cdot  \mathcal{L}_k )^{\ell+1}  }{(\hat q \cdot p_k)^{\ell+4} }\right]. 
    \end{split}
\end{equation}
With a strictly conformally covariant soft factor at hand, we can now proceed exactly as in the massless case.  In particular, the primary descendant is now guaranteed to be independent of the reference point and generalizes the simplification found in \eqref{pd-1} to $\ell>1$:
\begin{equation} \label{eq:massivedes}
    \begin{split}
	\partial^{\ell+3}_{\bar{z} }S^{(\ell)}_k(z, \bar{z})
	    = \mathcal{N}_\ell \frac{p_k^4}{(\hat q \cdot p_k)^{\ell+4}} \left(F_- \cdot \mathcal{L}_k \right)^{\ell+1}. 
    \end{split}
\end{equation}
In deriving \eqref{eq:massivedes}, it is helpful first to exchange powers of $F_+ \cdot \mathcal{L}_k$ for powers of $F_- \cdot \mathcal{L}_k$ by using the identity
\begin{equation} \label{eq:IdentityFpm-monica}
    \begin{aligned}
        (F_+ \cdot \mathcal{L}_k)^n=
       \frac{(\hat{q} \cdot p_k )^{2n+1}}{(p_k^2)^n (2n)!} \partial_{z}^{2n} \left( \frac{\left(F_- \cdot \mathcal{L}_k \right)^n}{\hat{q} \cdot p_k}\right),
    \end{aligned}
\end{equation}
which can be proved by induction. Combining this result with the result \eqref{pd-1} for $\ell = -1, 0, 1$, we arrive at the following set of transformations of massive particles, each now associated with a fixed order of the soft expansion:   
\begin{equation} \label{massive-tf}
    \begin{split}
        \delta^p_m |p_k \rangle &\equiv- \frac{1}{2}\int \frac{d^2z}{2 \pi } \bar{z}^{p+m-1} \partial_{\bar{z}}^{2p-1} \left \{ \begin{array}{cl}
                    S'^{(2p-4)}_k (z, \bar{z})| p_k \rangle,  
        &\quad p = \frac{3}{2},2, \frac{5}{2}\vspace{1ex}\\ 
                     S^{(2p-4)}_k (z, \bar{z})| p_k \rangle,  
        &\quad p > \frac{5}{2}\end{array}  \right. \\
        & =-\frac{  \mathcal{N}_{2p-4}}{2} \int \frac{d^2z}{2 \pi } \bar{z}^{p+m-1} \frac{p_k^4}{(\hat q \cdot   p_k)^{2p}} \left( F_- \cdot \mathcal{L}_k \right)^{2p-3}| p_k \rangle.
    \end{split}
\end{equation}
Note the last line applies for $p \geq \frac{3}{2}$, but omits the case when $p = 1$.  As observed in \cite{Strominger:2021lvk},  $p = 1$ corresponds to a central term in the wedge algebra.  No soft theorem analysis has yet provided a non-vanishing value for this central extension so we leave the question of an appropriate expression for massive scalars to a future investigation.  To complete the same analysis as for massless particles in the previous subsection, it remains to be shown that the transformations \eqref{massive-tf} generate a ${\rm w}_{1+\infty}$ symmetry action on massive particles. This is the subject of the next section. 

\section{Momentum space action of ${\rm w}_{1+\infty}$} \label{sec:momentumaction} 

In this section, we establish that the transformation 
\begin{equation} \label{massive-tf4}
    \begin{split}
        \delta^p_m |p_k \rangle &  =- \frac{  \mathcal{N}_{2p-4}}{2}\int \frac{d^2z}{2 \pi } \bar{z}^{p+m-1} \frac{p_k^4}{(\hat q \cdot   p_k)^{2p}} \left( F_- \cdot \mathcal{L}_k \right)^{2p-3}| p_k \rangle,
    \end{split}
\end{equation} 
respects the algebra
\begin{equation} \label{w-algebra4}
    \left[ \delta^p_m,  \delta^q_n\right]
	         = \left[m (q-1)-n(p-1)\right]  \delta^{p+q-2}_{m+n}.
\end{equation}
This result follows from an induction proof that mirrors the one presented in Appendix C of \cite{Himwich:2021dau}.  Here we will sketch the logic and point the reader to \cite{Himwich:2021dau} for explicit details omitted here. 
	
In \cite{Himwich:2021dau}, an abstract symmetry action $\delta^p_m$ with $p = 1, \frac{3}{2}, 2, \frac{5}{2}, \cdots$ and $1-p \leq m \leq p-1$ was proven generally to respect the algebra \eqref{w-algebra4}, provided that it respects base cases of \eqref{w-algebra4} with $(p,m) = (1,0)$, $(\frac{3}{2}, - \frac{1}{2})$, $(2,\pm1)$, $(2,0)$ and $(\frac{5}{2}, - \frac{3}{2})$ and $(q,n)$ compatible with \eqref{mode-range} but otherwise arbitrary.  The proof  in \cite{Himwich:2021dau} of \eqref{w-algebra4} for generic $p,m,q,n$ given these base cases depended solely on the abstract algebra \eqref{w-algebra4} and not on the explicit representation of $\delta^p_m$.  Hence we can directly import that result, thus reducing the general proof to establishing these base cases for the explicit transformation in \eqref{massive-tf4}.  Also note that we will not treat the case $(p,m)=(1,0)$ since our expression does not extend to this case, as discussed in the previous section.  Importantly, this base case was not used in proving the induction step in \cite{Himwich:2021dau} and thus its omission will not affect the above argument. 
	
First consider $(p,m) =  (\frac{3}{2}, - \frac{1}{2})$. Evaluating \eqref{massive-tf4}, we find 
\begin{equation} \label{eq:npaction}
    \begin{split}
	\delta^{\frac{3}{2}}_{-\frac{1}{2}}|p_k \rangle
	    & = \frac{\eps_k m_k}{4y_k} |p_k \rangle=  - \frac{1}{4} (n \cdot p_k) |p_k \rangle.
    \end{split} 
\end{equation}
To evaluate the commutator, we will need the identity 
\begin{equation}
    \begin{split}
	\left[ p_k^\mu , (F_- \cdot \mathcal{L}_k)^\ell \right]
	    = -2i\ell F_-^{\mu \nu} p_k{}_\nu (F_- \cdot \mathcal{L}_k)^{\ell-1}. 
    \end{split}
\end{equation}
Then, we find\footnote{Note to compute the commutators, we employ the standard interpretation of the ordering so that for example ${\rm SL}(2, \mathbb{C})$ transformations of the form $\delta^2_m \mathcal{O}  = \bar{z}^m \left[(m+1) \bar{h}+\bar{z} \partial_{\bar{z}} \right] \mathcal{O} $ obey the familiar commutation relation $\left[\delta^2_m, \delta^2_n\right] = (m-n) \delta^2_{m+n}$. This requires that
\begin{equation*}
    \delta^2_m \delta^2_n \mathcal{O}  = 
    \bar{z}^n \left((n+1) \bar{h}+\bar{z} \partial_{\bar{z}} \right)\delta^2_m   \mathcal{O} 
    =\bar{z}^n \left((n+1) \bar{h}+\bar{z} \partial_{\bar{z}} \right)\bar{z}^m \left((m+1) \bar{h}+\bar{z} \partial_{\bar{z}} \right)\mathcal{O} .
\end{equation*} \label{footnote} } 
\begin{equation}
    \begin{split}
	\left[ \delta^{\frac{3}{2}}_{-\frac{1}{2}}, \delta^q_n\right] |p_k\rangle
	    & =-\frac{1}{4} \frac{ \mathcal{N}_{2q-4}}{2} \int \frac{d^2z}{2 \pi } \bar{z}^{q+n-1} \frac{p_k^4}{(\hat q \cdot   p_k)^{2q}}  \left[n\cdot p_k, \left( F_- \cdot \mathcal{L}_k\right)^{2q-3} \right]|p_k\rangle\\
	    & =- \frac{1}{2} \frac{ \mathcal{N}_{2q-4}}{2}\int \frac{d^2z}{2 \pi } \bar{z}^{q+n-1} p_k^4~\frac{(-i)(2q-3)n_\mu p_{k\nu} F_-^{\mu\nu}}{(\hat q \cdot   p_k)^{2q}}     \left(F_- \cdot \mathcal{L}_k\right)^{2q-4} |p_k\rangle.
    \end{split}
\end{equation}
Next, we note that
\begin{equation}
    \begin{split}
	    -i\mathcal{N}_{2q-4} (2q-3)   \frac{n_\mu p_{k\nu} F_-^{\mu\nu}}{(\hat q \cdot   p_k)^{2q}}
	=\mathcal{N}_{2q-5}   \partial_{\bar{z}} \left(\frac{1}{(\hat q \cdot   p_k)^{2q-1}}\right).
    \end{split}
\end{equation}
Integrating by parts with respect to $\bar{z}$ and using the fact that $\partial_{\bar{z}}  \left(F_- \cdot \mathcal{L}_k\right) = 0$, we find  
\begin{equation} \label{eq:w32proof}
    \begin{split}
	\left[ \delta^{\frac{3}{2}}_{-\frac{1}{2}}, \delta^q_n\right] |p_k\rangle
	& =  \frac{1}{2}\left(q+n-1\right) \frac{ \mathcal{N}_{2q-5}}{2}  \int \frac{d^2z}{2 \pi } \bar{z}^{q+n-2} \frac{p_k^4}{(\hat q \cdot   p_k)^{2q-1}} \left(F_- \cdot \mathcal{L}_k\right)^{2q-4} |p_k\rangle\\
	& =  -\frac{1}{2} \left(q+n-1\right) \delta^{q-\frac{1}{2}}_{n-\frac{1}{2}}|p_k\rangle.
    \end{split}
\end{equation}

To establish \eqref{w-algebra4} when $p=2$ and $m = -1,0,1$, we again explicitly evaluate \eqref{massive-tf4} with these values: 
\begin{equation} \label{massive-w2}
    \begin{split}
        \delta^2_m |p_k \rangle &  = \underbrace{ \frac{1}{2} \bar{w}_k^{m-1}\left (2 \bar{w}_k^{2}\p_{\bar{w}_k} + (m+1)\bar{w}_k  y_k \p_{y_k} -m(m+1)  y_k^2 \p_{w_k} \right)}_{\equiv \bar{L}_m  }|p_k \rangle .
    \end{split}
\end{equation}
Notice these are just three of the Killing vectors on the unit hyperboloid parametrized by $(y_k, w_k, \bar{w}_k)$.  Hence, their commutators are just familiar ${\rm SL}(2, \mathbb{C})$ expressions.  We will need the following 
\begin{equation}
    \begin{split}
        \left[\bar{L}_m , \frac{1}{(\hat q \cdot   p_k)^{2q} }\right] & = -	\bz^m \left((m+1) q+ \bz \p_\bz \right) \frac{1}{(\hat q \cdot   p_k)^{2q} },\\
	\left[\bar{L}_m , F_- \cdot \mathcal{L}_k\right] &= 0.
    \end{split}
\end{equation}
Using these, we readily find 
\begin{equation}
    \begin{split}
        \left[\delta^2_m, \delta^q_n\right]|p_k \rangle
        & =\frac{\mathcal{N}_{2q-4}}{2} \int \frac{d^2z}{2 \pi} \bar{z}^{q+n-1} p_k^4
            \left[\bar{L}_m , \frac{(F_- \cdot \mathcal{L}_k)^{2q-3}}{(\hat q \cdot p_k)^{2q}} \right] |p_k\rangle\\
        & =-\frac{\mathcal{N}_{2q-4}}{2} \int \frac{d^2z}{2 \pi } \bar{z}^{q+n+m-1}  \left((m+1) q+ \bz \p_\bz \right)  p_k^4\frac{(F_- \cdot \mathcal{L}_k)^{2q-3}}{(\hat q \cdot p_k)^{2q} } |p_k\rangle.
    \end{split}
\end{equation}
Once again integrating by parts in $\bar{z}$, this becomes 
\begin{equation} \label{eq:delta2comm}
    \begin{split}
        \left[\delta^2_m, \delta^q_n\right]|p_k \rangle 
        & = -\left(m (q-1) -n \right)\frac{\mathcal{N}_{2q-4}}{2}\int \frac{d^2z}{2 \pi} \bar{z}^{q+n+m-1}(p_k)^4 \frac{(F_- \cdot \mathcal{L}_k)^{2q-3}}{(\hat q \cdot p_k)^{2q}}|p_k\rangle\\
        & = \left(m (q-1) -n \right) \delta^{q}_{m+n}|p_k \rangle .
    \end{split}
\end{equation}
    
Finally, we treat the case when $p = \frac{5}{2}$ and $m = - \frac{3}{2}$.  Proceeding as before, we evaluate \eqref{massive-tf4} for these values and find 
\begin{equation} 
    \begin{split}
        \delta^{\frac{5}{2}}_{-\frac{3}{2}} |p_k \rangle &  = \frac{ 3 y_k}{\eps_k m_k} \partial_{\bar{w}_k}^2| p_k \rangle.
    \end{split}
\end{equation}
Using this expression, one can verify by brute force that this generator likewise respects
\begin{equation} \label{eq:w52proof}
    \begin{split}
        \left[ \delta^{\frac{5}{2}}_{-\frac{3}{2}},  \delta^q_n\right] |p_k \rangle 
        &  =-\frac{3}{2} \left( q+n-1\right) \delta^{q+\frac{1}{2}}_{n-\frac{3}{2}}|p_k \rangle.
    \end{split}
\end{equation}
The details of this calculation are outlined in Appendix \ref{app:w52}.  Having thus established the base cases \eqref{eq:w32proof}, \eqref{eq:delta2comm}, and \eqref{eq:w52proof}, we can directly apply the induction proof in Appendix C of \cite{Himwich:2021dau} to establish that the symmetry transformation respects the algebra \eqref{w-algebra4} when acting on massive particles.
	
\section{Generating functions} \label{sec:genfun}
 
Before turning to the conformal primary basis, it is helpful to compare the results of our momentum-space analysis in the massless and massive cases. The ${\rm w}_{1+\infty}$ symmetry transformations of massless particles in the conformal primary basis can be easily deduced from momentum space expressions. In particular, the coordinates $z_k, \bar{z}_k$ are preserved and the energy scale $\omega_k$ raises the conformal weight $\Delta_k$ by one. By contrast, for massive particles, momentum becomes a non-trivial differential operator in the conformal primary basis.  Therefore, simple expressions for the charges like \eqref{massive-tf4} involving inverse powers of momentum must be treated with some care. 

To derive expressions in the conformal primary basis, we proceed in two steps.  First, we re-express the integral expression in sum form in this section. Second, in the next section, we transform to the conformal primary basis.  Rather than dealing with each mode $n$ of a transformation $\delta^q_n$ individually, we find it convenient to work with a generating function that neatly packages all the ${\rm SL}(2, \mathbb{C})$-related symmetry transformations associated to a particular order of the soft expansion.  We begin with a discussion of the generating function for massless particles. Previous analyses in $(2,2)$ signature \cite{Banerjee:2020kaa,Banerjee:2020zlg,Banerjee:2021cly,Banerjee:2021dlm,Banerjee:2020vnt,Guevara:2021abz} essentially treat the soft factor itself as a generating function.  In order to generalize to the massive case, we explain how the same massless generating function can be equivalently derived directly from the charges in $(1,3)$, as opposed to the soft factor in $(2,2)$.  Finally, we present the massive generalization and use the generating function to re-express the integral in \eqref{massive-tf4} in sum form, which will facilitate the transformation to the conformal primary basis in the next section.

\subsection{Massless particles} \label{sec:masslessgen}

We begin with the definition of the generators \eqref{def-wsym}. We can find a generating function by considering
\begin{equation}
  \begin{aligned}
    \mathbf{G}_p(\bar{a}) |p_k \rangle &\equiv \sum_{m = 1-p}^{p-1} \frac{\Gamma(2p-1)}{\Gamma(p+m)\Gamma(p-m)} (-\bar{a})^{p-m-1} \delta^p_m |p_k \rangle \\
    &=  -\frac{1}{2}\int \frac{d^2z}{2 \pi } \left(\bar{z}- \bar{a}\right)^{2p-2} \partial_{\bar{z}}^{2p-1} S'^{(2p-4)}_k (z, \bar{z}) |p_k \rangle. 
  \end{aligned}
\end{equation}
Now, integrating by parts, using the explicit form of the massless soft factor primary descendant \eqref{masslesspd}, the form of $F_+ \cdot \mathcal{L}_k$ \eqref{eq:FLmassless} and $\partial_z \hat{q}\cdot p_k$ for massless particles, and relabelling $a \to z$, we find 
\begin{equation} \label{eq:masslessgenerating}
  \begin{aligned}
   \mathbf{G}_p(\bar{z})|p_k \rangle 
   &=  \frac{1}{2} (2p-2) (-1)^{2p-3}\epsilon_k\omega_k(\bar{z}_k - \bar{z})  \left(\frac{1}{\epsilon_k\omega_k} \left[\omega_k \partial_{\omega_k} - (\bar{z}_k - \bar{z})\partial_{\bar{z}_k}\right]\right)^{2p-3} |p_k \rangle\\
   &=  \frac{\mathcal{N}_{2p-4} }{2(2p-1)} \partial_z \hat{q} \cdot p_k \left(\frac{F_+ \cdot \mathcal{L}_k}{\partial_z \hat{q} \cdot p_k}\right)^{2p-3}|p_k \rangle. 
  \end{aligned}
\end{equation}
Note that this expression is a polynomial in positive powers of $(\bar{z} - \bar{z}_k)$. The powers of $ F_+ \cdot \mathcal{L}_k$  naturally produce the  expected action in terms of $\bar{z}$ descendants. Although the same powers of $F_+ \cdot \mathcal{L}_k$ appear in the soft factor \eqref{softfactor1} and the right-hand side of \eqref{eq:masslessgenerating}, the two expressions fundamentally differ because the right-hand side of \eqref{eq:masslessgenerating} is independent of the reference point $z_0$ while the soft factor (for $p > \frac{5}{2}$) is not, as emphasized in Section \ref{sec:massless}.   

This same generating function equivalently can be found by taking a contour integral of the massless soft factor, which is a similar prescription to that in a standard 2D CFT and to that in \cite{Himwich:2021dau} for determining the ${\rm w}_{1+\infty}$ symmetry action on massless particles. In particular, when $z$ and $\bar{z}$ are regarded as independent complex variables, the contour integral of the soft factor \eqref{softfactor1} reproduces \eqref{eq:masslessgenerating} up to normalization: 
\begin{equation}  \label{eq:soft1contour}
  \oint_{z_k} \frac{dz}{2\pi i} S'^{(2p-4)}_k (z, \bar{z})|p_k \rangle =  \frac{(-1)^{2p}\mathcal{N}_{2p-4}}{(2p-1)!} \partial_z \hat{q} \cdot p_k \left(\frac{F_+ \cdot \mathcal{L}_k}{\partial_z \hat{q} \cdot p_k}\right)^{2p-3}|p_k \rangle.
\end{equation}
Note that the reference point $z_0$ is also projected out by the contour integral. Recognizing that \eqref{eq:soft1contour} is related to the sum form of $\mathbf{G}_p(\bar{z})$ by \eqref{eq:masslessgenerating}, we recover precisely the form of the discrete light transform that appears in \cite{Himwich:2021dau,Strominger:2021lvk}:
\begin{equation}
  \begin{aligned}
    \oint_{z_k} \frac{dz}{2\pi i}  S'^{(2p-4)}_k (z, \bar{z}) |p_k \rangle= \sum_{m = 1-p}^{p-1} \frac{(-1)^{2p}}{\frac{1}{2}\Gamma(p+m)\Gamma(p-m)} (-\bar{z})^{p-m-1} \delta^p_m |p_k \rangle .
  \end{aligned}
\end{equation}
The coefficients in the sum are precisely the ones that were previously introduced by hand in \cite{Himwich:2021dau,Strominger:2021lvk} to  make the ${\rm w}_{1+\infty}$ symmetry manifest. Here we see that they appear naturally in the expansion of the contour integral of the soft factor in powers of $\bar{z}$.

\subsection{Massive particles} \label{sec:massivegen}

In this subsection, we present a generating function for the momentum-space action of ${\rm w}_{1+\infty}$ on massive particles,  which can be directly compared with the generating function for the massless case in the previous subsection. 

In the massless case, the generators involve descendants of the soft factor \eqref{masslesspd} that can be naturally rewritten in terms of $ F_+ \cdot \mathcal{L}_k$, as we saw in the previous subsection. In the massive case, however, we begin from an integral expression involving angular momentum generators of the opposite helicity. To write a generating function in a form analogous with that of the massless case, we proceed in two steps: first, find a sum (rather than integral) expression for the generating function for the generators \eqref{massive-tf4}, and second, rewrite that expression in terms of the right-moving angular momentum generators. 

First, in analogy with the massless case, we consider the integral 
\begin{equation} \label{eq:massivegenstep}
 \mathbf{G}_p(\bar{a}) |p_k \rangle \equiv  - \frac{\mathcal{N}_{2p-4}}{2}\int \frac{d^2z}{2 \pi } (\bar{z}-\bar{a})^{2p-2}  \frac{p_k^4}{(\hat q \cdot   p_k)^{2p}} \left( F_- \cdot \mathcal{L}_k \right)^{2p-3} |p_k \rangle. 
\end{equation}
To turn this integral into a sum, we recognize that it can be recast in terms of a shadow transform \cite{Simmons-Duffin:2012juh} (recall that $\hat{q} \cdot p_k$ has weight $(h,\bar{h}) = \left(-\frac{1}{2},-\frac{1}{2}\right)$):
\begin{equation} \label{eq:shadowtrans}
\int \frac{d^2 z}{2\pi} \frac{\left[(z-a)(\bar{z}-\bar{a})\right]^{2p-2}}{\left(\hat{q}\cdot p_k\right)^{2p}} = \mathcal{C}_p\frac{\left(\hat{q}(a) \cdot p_k\right)^{2p-2}}{m_k^{4p-2}},
\end{equation}
where $\mathcal{C}_p$ is a normalization given by 
\begin{equation}
\mathcal{C}_{p} = \frac{\Gamma(2p-1)}{\Gamma(2p)} = \frac{1}{(2p-1)}.
\end{equation}
Expanding $ F_- \cdot \mathcal{L}_k$ in powers of $(z-a)$, we have
\begin{equation}
    \begin{split}
         \mathbf{G}_p(\bar{a}) |p_k \rangle= - \frac{  \mathcal{N}_{2p-4}}{2} \sum_{n = 0}^{4p-6} \frac{1}{n!} \left( \int \frac{d^2z}{2 \pi } (\bar{z}-\bar{a})^{2p-2} (z-a)^n \frac{p_k^4}{(\hat q \cdot   p_k)^{2p}} \right)
            \partial_a^n\left( F_-(a) \cdot \mathcal{L}_k \right)^{2p-3} |p_k \rangle. 
    \end{split}
\end{equation}
Then using
\begin{equation}
    \begin{split}
        (-\partial_a)^B (z-a)^A = \frac{A!}{(A-B)!} (z-a)^{A-B},
    \end{split}
\end{equation}
which remains formally true for inverse derivatives interpreted as definite integrals from $z$ to $a$, we find
\begin{equation}
    \begin{split}
         \mathbf{G}_p(\bar{a}) |p_k \rangle = - \frac{  \mathcal{N}_{2p-4}}{2 (2p-2)!} \sum_{n = 0}^{4p-6} \Big( (-\partial_a)^{2p-2-n} \int \frac{d^2z}{2 \pi } (\bar{z}-\bar{a})^{2p-2} &(z-a)^{2p-2} \frac{p_k^4}{(\hat q \cdot   p_k)^{2p}} \Big) \\
         &\times \partial_a^n\left( F_-(a) \cdot \mathcal{L}_k \right)^{2p-3} |p_k \rangle. 
    \end{split}
\end{equation}
Finally, using the shadow transform \eqref{eq:shadowtrans} and relabelling $a\to z$, we have 
\begin{equation} \label{eq:massivemidstep}
\begin{aligned}
 \mathbf{G}_p(\bar{z}) |p_k \rangle= - \frac{  \mathcal{N}_{2p-4}\mathcal{C}_p}{2 (2p-2)! m_k^{4p-6}} \sum_{n = 0}^{4p-6}  (-\partial_z)^{2p-2-n}  (\hat q   \cdot p_k)^{2p-2} ~
            \partial_z^n\left( F_-(z) \cdot \mathcal{L}_k \right)^{2p-3} |p_k \rangle.
 \end{aligned}
\end{equation}
To compare this expression to the massless case, and also to rewrite the generating function in an explicit form that does not involve formal inverse powers of derivatives, we use the identity
\begin{equation} \label{eq:Fpm}
    \begin{aligned}
        ( F_- \cdot \mathcal{L}_k)^n = \frac{ (-1)^n}{(2n)! m_k^{2n}} \sum_{m = 0}^{2n} \left[(-\partial_{\bar{z}})^{2n-m} (p_k \cdot \hat q )^{2n}\right]~ \partial_{\bar{z}}^{m}\left( F_+ \cdot \mathcal{L}_k \right)^n, 
    \end{aligned}
\end{equation}
which follows from the identity \eqref{eq:IdentityFpm-monica}. Now, using the completeness relation \eqref{completeness1} for massive particles to expand the factors of $\hat{q} \cdot p_k$ and evaluating the derivatives, we ultimately find
\begin{equation} \label{eq:massivegenfun}
  \begin{aligned}
     \mathbf{G}_p(\bar{z}) |p_k \rangle = \frac{\mathcal{N}_{2p-4}\mathcal{C}_p}{2\left(-n \cdot p_k\right)^{2p-4}}  \sum_{\ell=0}^{4p-6} \frac{1}{\ell!} \left(\sum_{j=2p-4}^{4p-6} (-1)^j {\ell \choose j} \right)\left(- \frac{\partial_z \hat{q}\cdot p_k}{n\cdot p_k} \right)^{\ell - 2p + 4} \partial_{\bar{z}}^{\ell}  \left( F_+ \cdot \mathcal{L}_k\right)^{2p-3} |p_k \rangle. 
  \end{aligned}
\end{equation}
More details of the derivation of this result are presented in Appendix \eqref{app:massivegen}.  Note that this expression is a polynomial in positive powers of  $(\bar{z} - \bar{w}_k)$. For particular values of $p$, the generating functions become
\begin{equation} \label{eq:massivegenspec}
  \begin{aligned}
    \mathbf{G}_{\frac{3}{2}}(\bar{z})|p_k \rangle &= \frac{1}{4} \partial_z\hat{q} \cdot p_k |p_k \rangle  , \\
    \mathbf{G}_2(\bar{z}) |p_k \rangle&= -\frac{\sqrt{2}}{2} \left( F_+ \cdot \mathcal{L}_k\right)|p_k \rangle, \\
    \mathbf{G}_{p>2}(\bar{z})|p_k \rangle
    &=  \frac{\mathcal{N}_{2p-4}\mathcal{C}_p}{2\left(-n \cdot p_k\right)^{2p-4}} \frac{(-1)^{2p}}{(2p{-}5)!} \sum_{\ell=2p-4}^{4p-6} \frac{1}{\ell(\ell{-}2p{+}4)!} \left(- \frac{\partial_z \hat{q}\cdot p_k}{n\cdot p_k} \right)^{\ell{-}2p{+}4} \partial_{\bar{z}}^{\ell}  \left( F_+ \cdot \mathcal{L}_k\right)^{2p-3}|p_k \rangle.
  \end{aligned}
\end{equation}
For $p=\frac{3}{2}$ and $p=2$ these exactly match the expressions found in the previous subsection. For $p >2$, the massless limit of these expressions reduces to the form found in the previous subsection. For massive momenta when $p>2$, \eqref{eq:massivegenspec} is not equivalent to the massless expression \eqref{eq:masslessgenerating} because the massive soft factor was modified as described in Section \ref{sec:massive}. This modification was ultimately required to obtain conformally covariant expressions.

\section{Action of ${\rm w}_{1+\infty}$ on massive celestial conformal primaries} \label{sec:conformalpb}

In this section, we formulate the action of ${\rm w}_{1+\infty}$ on massive scalar celestial primaries by transforming the results of the previous section to the conformal primary basis.

Although we now have expressions for the symmetry action in terms of simple powers of $\hat{q} \cdot p_k$, $n \cdot p_k$, and $F_+ \cdot \mathcal{L}_k$, one subtlety still remains in determining the associated action on massive particles in the conformal primary basis: the momentum-space expressions for the charges involve inverse powers of the four-momentum $p_k$.  In the conformal primary basis, the four-momentum $P$ becomes a non-trivial differential operator \cite{Law:2019glh,Law:2020tsg} that acts on massive scalars as
\begin{equation} \label{zl-generators}
    \begin{split}
        P_k^{\mu}  \Phi_{\Delta_k}&(z_k, \bar{z}_k)
        \\& = \frac{\eps_k m_k}{2} \Bigg[ \left(\partial_{z_k} \partial_{\bar{z}_k} \hat q^{\mu}_k
            + \frac{\left(\partial_{\bar{z}_k} \hat q^{\mu}_k\right) \partial_{z_k}+\left(\partial_{z_k} \hat q^{\mu}_k\right) \partial_{\bar{z}_k} }{\Delta_k -1}  + \frac{\hat q_k^\mu \partial_{z_k} \partial_{\bar{z}_k}}{(\Delta_k -1)^2}\right)     e^{-\partial_{\Delta_k}}\\
            & \hspace{57ex}+   \frac{\Delta_k \hat q^\mu_k}{\Delta_k -1}e^{\partial_{\Delta_k}}\Bigg] \Phi_{\Delta_k}(z_k,  \bar{z}_k),
    \end{split}
\end{equation}
so in the case of massive external particles, inverse powers of momentum must be treated with some care. This is in contrast to the massless conformal primary basis, in which the action of four-momentum is simply weight-raising and so inverse powers of momentum can be easily interpreted as weight-lowering. 

One benefit of the generating function form \eqref{eq:massivegenfun} as a sum, rather than an integral, is that it makes manifest the fact that inverse powers of the differential operator $P_k$,  as opposed to just the scalar multiplier $m_k$, are necessarily involved in the expressions for the massive generators for $p > 2$.  To work with these expressions in the conformal primary basis, it is therefore helpful to provide a prescription for performing calculations.

A natural (albeit formal) interpretation of the inverse powers is to replace them with a Schwinger parametrization: specifically, recalling the action \eqref{eq:npaction} of $n\cdot p_k$, consider making the replacement
\begin{equation} \label{replacement}
  \frac{4^m}{(- n\cdot p_k)^m} |p_k\rangle  \mapsto \left(\delta^{\frac{3}{2}}_{-\frac{1}{2}}\right)^{-m} |p_k\rangle \equiv \frac{1}{ (m-1)!} \int_0^{\infty} dx \ x^{m-1}\sum_{j=0}^{\infty} \frac{1}{j!}(-x \ \delta^{\frac{3}{2}}_{-\frac{1}{2}})^j |p_k\rangle.
\end{equation}
The Schwinger parametrization is  a popular trick for working with formal expressions involving inverted differential operators and has even appeared before in studies of the matrix elements of momentum generators in a Lorentz basis \cite{MacDowell1972}. To prove that  the parametrization of the symmetry action with \eqref{replacement} will give the same algebra, we must simply check that this replacement gives the same commutators with the $\delta^{\frac{3}{2}}_n$ and $\delta^{2}_n$ generators, because $\hat{q}\cdot p_k$ and $F_+ \cdot \mathcal{L}_k$ are constructed from these modes and all of the generators in \eqref{eq:massivegenfun} are constructed from powers of these operators. Since the momenta commute, the only nontrivial cases to check are $\delta^{2}_{0}$ and $\delta^{2}_{1}$. Before making the replacement \eqref{replacement}, we compute the commutators explicitly from their derivative form \eqref{massive-w2}: 
\begin{equation}
  \begin{aligned}
    \left[\frac{1}{2}y_k\partial_{y_k} + \bar{w}_k\partial_{\bar{w}_k},\left(\frac{y_k}{\epsilon_k m_k}\right)^m\right] |p_k\rangle &= \frac{m}{2}\left(\frac{y_k}{\epsilon_k m_k}\right)^m |p_k\rangle, \\
    \left[\bar{w}_k^2\partial_{\bar{w}_k} + \bar{w}_ky_k\partial_{y_k} - y_k^2 \partial_{w_k},\left(\frac{y_k}{\epsilon_k m_k}\right)^m\right] |p_k\rangle &= m \bar{w}_k\left(\frac{y_k}{\epsilon_k m_k}\right)^m = m \left(\frac{\bar{w}_k \epsilon_k m_k}{y_k}\right)\left(\frac{y_k}{\epsilon_k m_k}\right)^{m+1} |p_k\rangle.
  \end{aligned}
\end{equation}
After making the replacement, it is straightforward to check using the commutators $\left[\delta^2_n,\delta^{\frac{3}{2}}_{-\frac{1}{2}}\right]$ and the expression in \eqref{replacement} that  
\begin{equation}
  \begin{aligned}
    \left[\delta^{2}_0,\left(\delta^{\frac{3}{2}}_{-\frac{1}{2}}\right)^{-m}\right] |p_k\rangle &= - \frac{m}{2}\left(\delta^{\frac{3}{2}}_{-\frac{1}{2}}\right)^{-m} |p_k\rangle, \\
    \left[\delta^{2}_1,\left(\delta^{\frac{3}{2}}_{-\frac{1}{2}}\right)^{-m}\right] |p_k\rangle &= - m\delta^{\frac{3}{2}}_{\frac{1}{2}} \left(\delta^{\frac{3}{2}}_{-\frac{1}{2}}\right)^{-m-1} |p_k\rangle.
  \end{aligned}
\end{equation}
Recalling the sign difference in derivative ordering (see footnote \ref{footnote}), the commutators before and after the replacement precisely match and confirm that $(\delta^{\frac{3}{2}}_{-\frac{1}{2}})^{-m}$ has right-weight $\frac{m}{2}$ and that $\delta^2_1$ has the proper raising action. Thus, to transform to the conformal primary basis, we simply use the replacement
\begin{equation}
  \bar{w}_k = - \frac{\partial_z\hat{q}\cdot p_k |_{\bar{z} = 0}}{n \cdot p_k} \mapsto \delta^{\frac{3}{2}}_{\frac{1}{2}}\left(\delta^{\frac{3}{2}}_{-\frac{1}{2}}\right)^{-1}, \ \ \ \frac{1}{-n\cdot p_k} \mapsto \frac{1}{4}\left(\delta^{\frac{3}{2}}_{-\frac{1}{2}}\right)^{-1}. 
\end{equation}

With this new interpretation of the generators in \eqref{eq:massivegenfun} involving positive powers of momentum, the action in the conformal primary basis now readily follows from the action in momentum space:
\begin{equation} \label{cpb-act1}
    \begin{split}
	\delta^q_n \Phi_{\Delta_k}(z_k, \bar{z}_k) \equiv  
	    \int [d \hat p_k ] ~ \mathcal{G}_{\Delta_k}(\hat p_k; \hat q_k)   \delta^q_n | p_k \rangle.
    \end{split}
\end{equation}
To evaluate this action, first note that the bulk-to-boundary propagator satisfies 
\begin{equation} \label{zl-bb}
    \begin{split}
        p_k^{\mu} \mathcal{G}_{\Delta_k}&(\hat p_k; \hat q_k) 
             \\& = \frac{\eps_k m_k}{2} \Bigg[ \left(\partial_{z_k} \partial_{\bar{z}_k} \hat q^{\mu}_k
            + \frac{\left(\partial_{\bar{z}_k} \hat q^{\mu}_k\right) \partial_{z_k}+\left(\partial_{z_k} \hat q^{\mu}_k\right) \partial_{\bar{z}_k} }{\Delta_k -1}  + \frac{\hat q_k^\mu \partial_{z_k} \partial_{\bar{z}_k}}{(\Delta_k -1)^2}\right)     e^{-\partial_{\Delta_k}}\\
            & \hspace{57ex}+   \frac{\Delta_k \hat q^\mu_k}{\Delta_k -1}e^{\partial_{\Delta_k}}\Bigg] \mathcal{G}_{\Delta_k}(\hat p_k; \hat q_k) ,
    \end{split}
\end{equation}
(which is how one verifies \eqref{zl-generators}). Collecting the transformations \eqref{cpb-act1} in the generating function from the previous section, we find
\begin{equation} \label{cpb-act2}
  \begin{aligned}
    \mathbf{G}_p(\bar{z}) \Phi_{\Delta_k}(z_k, \bar{z}_k)
    = \frac{\mathcal{N}_{2p-4}\mathcal{C}_p}{2\left(-n \cdot P_k\right)^{2p-4}}\sum_{\ell=0}^{4p-6}& \frac{1}{\ell!} \left(\sum_{j=2p-4}^{4p-6} (-1)^j {\ell \choose j} \right)\left(- \frac{\partial_z \hat{q}\cdot P_k}{n\cdot P_k} \right)^{\ell - 2p + 4} \\
    & \times  \partial_{\bar{z}}^{\ell}  \int [d \hat p_k ] ~ \mathcal{G}_{\Delta_k}(\hat p_k; \hat q_k)  \left(F_+ \cdot \mathcal{L}_k\right)^{2p-3} |p_k\rangle,
  \end{aligned}
\end{equation}
where in the above expression capital $P_k$ is used to indicate that momentum is now the operator appearing in \eqref{zl-generators} and inverse powers are evaluated using the Schwinger parametrization as in \eqref{replacement}. 
	
Next, note that the bulk-to-boundary propagator is an intertwiner and thus obeys
\begin{equation} \label{eq:Lact}
    \begin{split}
	\left( F_+ \cdot \mathcal{L}_k \right)\mathcal{G}_{\Delta_k}(\hat p_k; \hat q_k) 
	= \sqrt{2}   \left[ \Delta_k (\bar{z}_k-\bar{z}) + (\bar{z}_k-\bar{z})^2 \partial_{\bar{z}_k}\right]\mathcal{G}_{\Delta_k}(\hat p_k; \hat q_k). 
    \end{split}
\end{equation}
As a result, we can integrate by parts and then further pull this action outside the momentum integral to find 
\begin{equation} \label{cpb-act3}
    \begin{split}
	\mathbf{G}_p(\bar{z}) \Phi_{\Delta_k}(z_k, \bar{z}_k)
    = \frac{\mathcal{N}_{2p-4}\mathcal{C}_p}{2\left(-n \cdot P_k\right)^{2p-4}} \sum_{\ell=0}^{4p-6}& \frac{1}{\ell!} \left(\sum_{j=2p-4}^{4p-6} (-1)^j {\ell \choose j} \right)\left(- \frac{\partial_z \hat{q}\cdot P_k}{n\cdot P_k} \right)^{\ell - 2p + 4}\\&  
 \times \partial_{\bar{z}}^{\ell}  \left(-\sqrt{2}   \left[ \Delta_k (\bar{z}_k-\bar{z}) + (\bar{z}_k-\bar{z})^2 \partial_{\bar{z}_k}\right]\right)^{2p-3}  \Phi_{\Delta_k}(z_k, \bar{z}_k) ,
    \end{split}
\end{equation}
where we have identified
\begin{equation}
    \begin{split}
	\int [d \hat p_k ] ~ \mathcal{G}_{\Delta_k}(\hat p_k; \hat q_k)  ~   |p_k\rangle = \Phi_{\Delta_k}(z_k, \bar{z}_k).
    \end{split}
\end{equation}
	
As in the massless case, we identify the leading $q = \frac{3}{2}$ generators as simply reproducing the action of translations \eqref{zl-generators} in the conformal primary basis. Likewise, the $q = 2$ charges generate the action of (half of) ${\rm SL}(2, \mathbb{C})$.  Explicitly, 
\begin{equation}
    \begin{split}
        \delta^2_m \Phi_{\Delta_k}(z_k, \bar{z}_k)
        &= 	\bar{z}_k^m \left((m+1)\frac{\Delta_k}{2} + \bar{z}_k \p_{\bar{z}_k} \right)\Phi_{\Delta_k}(z_k, \bar{z}_k)  \\
        &= \bar{L}_m\Phi_{\Delta_k}(z_k, \bar{z}_k).
    \end{split}
\end{equation}

When $p>2$, the inverse powers of $n\cdot p_k$ in the momentum-space generating function become an infinite number of weight-shifting operators in the conformal primary basis, as is made explicit by the Schwinger parameter representation.  As such, the action of ${\rm w}_{1+\infty}$ on massive celestial primaries is notably more complicated than on massless celestial primaries, where the action maps a single conformal family to a single other conformal family. By contrast, in the massive case, a single conformal family is mapped by generators with $p>2$ to an infinite number of other conformal families. It would interesting to investigate whether some form of simplification or organizing structure materializes when working with a discrete conformal primary basis involving only integer-valued conformal dimensions. 

\section*{Acknowledgements}
		
 We are grateful to Nick Agia, Alfredo Guevara, Hofie Hannesdottir, Aidan Herderschee, Y.T. Albert Law, Donal O'Connell, and Ana Raclariu for insightful conversations. This work was supported by NSF grant 2310633.

\begin{appendix}
	
\section{Non-local behavior in conformally soft theorems}
\label{four-point-section}
		    
Consider a $U(1)$ gauge theory with three species of charged scalar particles, $\phi_1$, $\phi_3$ and $\Phi_4$, which couple via the three-point interaction 
\begin{equation}
    \begin{split}
	\mathcal{L}_{\rm int} \sim g \phi_1 \phi_3 \Phi_4.
    \end{split}
\end{equation}
Let $\phi_1$ and $\phi_3$ be massless and $\Phi_4$ be massive with mass $m$. Their respective charges, $Q_1$, $Q_3$, and $Q_4$, are required by the $U(1)$ gauge symmetry to obey charge conservation:
\begin{equation}
    \begin{split}
        Q_1+Q_3+Q_4 = 0.
    \end{split}
\end{equation}
In this appendix, we will see that a contour integral of a 2D $U(1)$ current encircling the operator insertions will vanish, as we expect in a standard conformal field theory.  However, unlike in a standard conformal field theory, there are non-local contributions to the contour integral that are due to the presence of a massive particle.

\subsection{Three-point interaction}
		     
Here we review the celestial three-point amplitude coupling one massive and two massless charged particles.  Note that this amplitude has previously appeared in \cite{Lam:2017ofc}. The three-point celestial amplitude takes the general form
\begin{equation} \label{3-point-gen}
    \begin{split}
        \langle \phi_{\Delta_1}  (z_1, \bar{z}_1)&  \phi_{\Delta_3}  (z_3, \bar{z}_3) \Phi_{\Delta_4} (z_4, \bar{z}_4)\rangle\\
	& = \int_0^\infty \frac{d \omega_1}{\omega_1} ~ \omega_1^{\Delta_1}
        \int_0^\infty \frac{d \omega_3}{\omega_3} ~ \omega_3^{\Delta_3}\int [d \hat p_4]~\mathcal{G}_{\Delta_4} (\hat p_4; \hat q_4) ~\mathcal {A}_3 (  \eps_1 \omega_1 \hat q_1, \eps_3 \omega_3 \hat q_3,\eps_4  m \hat p_4),
    \end{split}
\end{equation}
where $\mathcal{A}_3$ is the momentum-space three-point amplitude
\begin{equation}
    \mathcal {A}_3 (p_1, p_2, p_3) = g \delta^{(4)}(p_1+p_2+p_3)
\end{equation}
and 
\begin{equation}
    \hat q_k^\mu \equiv \hat q^\mu (z_k, \bar{z}_k).
\end{equation}
Using ${\rm SL}(2, \mathbb{C})$ covariance to send $z_1 \to 0$, $z_3 \to 1$ and $z_4 \to \infty$, \eqref{3-point-gen} becomes
\begin{equation} \label{3-point-simp}
    \begin{split}
        \langle \phi_{\Delta_1}  (z_1, \bar{z}_1)  \phi_{\Delta_3}  (z_3, \bar{z}_3) \Phi_{\Delta_4} (z_4, \bar{z}_4) \rangle
        = \frac{\lim_{\tau, \bar{\tau} \to \infty} \tau^{\Delta_4} \bar{\tau}^{\Delta_4} \langle \phi_{\Delta_1}  (0,0)  \phi_{\Delta_3}  (1,1) \Phi_{\Delta_4}  (\tau, \bar{\tau}) \rangle}
        {(z_{41}\bar{z}_{41})^{\frac{1}{2} (\Delta_1+\Delta_4-\Delta_3 )} 
        (z_{13}\bar{z}_{13})^{\frac{1}{2} (\Delta_1+\Delta_3-\Delta_4 )} 
        (z_{34}\bar{z}_{34})^{\frac{1}{2} (\Delta_3+\Delta_4-\Delta_1 )} } ,
    \end{split}
\end{equation}
where
\begin{equation} \label{3-point-eval}
    \begin{split}
	\lim_{\tau, \bar{\tau} \to \infty}  \tau^{\Delta_4} \bar{\tau}^{\Delta_4}  & \langle \phi_{\Delta_1}  (0,0)  \phi_{\Delta_3}  (1,1) \Phi_{\Delta_4}  (\tau, \bar{\tau}) \rangle\\
	&=  \int_0^\infty \frac{d \omega_1}{\omega_1} ~ \omega_1^{\Delta_1}
		\int_0^\infty \frac{d \omega_3}{\omega_3} ~ \omega_3^{\Delta_3} \int_0^\infty \frac{d y_4}{y_4^3} \int d^2 w_4 ~ y_4^{\Delta_4}\\& \quad \quad \times 
		g \delta^{(4)} \left(  \eps_1 \omega_1 \hat q (0,0)+ \eps_3 \omega_3 \hat q (1,1)+ \eps_4 \frac{m }{2y_4} (y_4^2 n  + \hat q  (w_4, \bar{w}_4) )\right).
    \end{split}
\end{equation} 
The solution to the delta function constraint is 
\begin{equation} \label{delta-solution}
	w_4 = \bar{w}_4 = \frac{m^2}{m^2 + 4 \omega_1^2}, \quad \quad \quad y_4 =- \eps_1 \eps_4\frac{2m \omega_1}{m^2 + 4 \omega_1^2}, \quad \quad \quad \omega_3 = \eps_1 \eps_3 \frac{m^2}{4 \omega_1}.
\end{equation}
Since both $y_4$ and $\omega_3$ are positive, notice that scattering is only nontrivial when $\eps_1 = \eps_3 = -\eps_4$.  Assuming this, the delta function can be written as
\begin{equation}  \label{delta-identity}
    \begin{split}
	\delta^{(4)} &\left( \eps_1 \omega_1 \hat q (0,0)+ \eps_3 \omega_3 \hat q (1,1)+ \eps_4 \frac{m }{2y_4} (y_4^2 n  + \hat q  (w_4, \bar{w}_4) )\right) \\&
	\quad \quad \quad \quad \quad  = \frac{8m \omega_1^2}{(m^2 + 4 \omega_1^2)^3} \delta^{(2)} \left(w_4-\frac{m^2}{m^2 + 4 \omega_1^2}\right) \delta \left(y_4-\frac{2m \omega_1}{m^2 + 4 \omega_1^2}\right) \delta\left(\omega_3 -\frac{m^2}{4 \omega_1} \right) .
    \end{split}
\end{equation}
Using this result to perform the integrals, we find 
\begin{equation}
    \begin{split}
	\lim_{\tau, \bar{\tau} \to \infty}  \tau^{\Delta_4} \bar{\tau}^{\Delta_4}  & \langle \phi_{\Delta_1}  (0,0)  \phi_{\Delta_3}  (1,1) \Phi_{\Delta_4}  (\tau, \bar{\tau}) \rangle = \frac{g}{8} \left(\frac{m}{2}\right)^{\Delta_1+\Delta_3-4}  B \left(\frac{\Delta_4+\Delta_1-\Delta_3}{2},\frac{\Delta_4-\Delta_1+\Delta_3}{2}\right) .
    \end{split}
\end{equation}
Substituting this into \eqref{3-point-simp}, we obtain the following closed form expression for the three-point celestial amplitude:
\begin{equation} \label{3-point-final}
    \begin{split}
	\langle \phi_{\Delta_1}  (z_1, \bar{z}_1)  \phi_{\Delta_3}  (z_3, \bar{z}_3) \Phi_{\Delta_4} (z_4, \bar{z}_4) \rangle
	= \frac{ \frac{g}{8} \left(\frac{m}{2}\right)^{\Delta_1+\Delta_3-4}     B \left(\frac{\Delta_4+\Delta_1-\Delta_3}{2},\frac{\Delta_4-\Delta_1+\Delta_3}{2}\right)}
	{(z_{41}\bar{z}_{41})^{\frac{1}{2} (\Delta_1+\Delta_4-\Delta_3 )} 
	(z_{13}\bar{z}_{13})^{\frac{1}{2} (\Delta_1+\Delta_3-\Delta_4 )} 
	(z_{34}\bar{z}_{34})^{\frac{1}{2} (\Delta_3+\Delta_4-\Delta_1 )} } .
    \end{split}
\end{equation} 
			
\subsection{Four-point conformally soft theorem}
	    
Now consider the celestial amplitude for the emission of a conformally soft photon $J$. Once again, we use conformal invariance to write this as 
\begin{equation} \label{4-point-gen}
    \begin{split}
	\langle \phi_{\Delta_1}(z_1, \bar{z}_1) &
      J(z_2,\bar{z}_2 )\phi_{\Delta_3}  (z_3, \bar{z}_3) \Phi_{\Delta_4}   (z_4, \bar{z}_4)\rangle \\	
      & =  \frac{z_{41}z_{34}}{z_{13} z_{24}^2}
            \frac{\lim_{\tau, \bar{\tau} \to \infty}  \tau^{\Delta_4} \bar{\tau}^{\Delta_4} 
            \langle   \phi_{\Delta_1}  (0,0) 
                J(z,\bar{z})\phi_{\Delta_3}  (1,1) \Phi_{\Delta_4}   (\tau,\bar{\tau})\rangle}{(z_{41}\bar{z}_{41})^{\frac{1}{2} (\Delta_1+\Delta_4-\Delta_3 )} 		(z_{13}\bar{z}_{13})^{\frac{1}{2} (\Delta_1+\Delta_3-\Delta_4 )} 	(z_{34}\bar{z}_{34})^{\frac{1}{2} (\Delta_3+\Delta_4-\Delta_1 )} } .
    \end{split}
\end{equation}
Here
\begin{equation}
    z =  \frac{z_{12}  z_{34}}{z_{13} z_{24}} 
\end{equation}
is the conformal cross ratio and 
\begin{equation} \label{4-point-red}
    \begin{split}
	&\lim_{\tau, \bar{\tau} \to \infty}  \tau^{\Delta_4} \bar{\tau}^{\Delta_4} 
        \langle \phi_{\Delta_1}(0,0) J(z,\bar{z})
        \phi_{\Delta_3}(1,1) \Phi_{\Delta_4}(\tau,\bar{\tau})\rangle\\
	& \quad  = \sqrt{2} \int_0^\infty \frac{d \omega_1}{\omega_1} ~ \omega_1^{\Delta_1}
			\int_0^\infty \frac{d \omega_3}{\omega_3} ~ \omega_3^{\Delta_3} \int_0^\infty \frac{d y_4}{y_4^3} \int d^2 w_4 ~ y_4^{\Delta_4} \mathcal{A}_4( \eps_1 \omega_1 \hat q (0,0), \hat q(z, \bar{z}), \eps_3 \omega_3 \hat q(1,1), \eps_4 m \hat p_4).
    \end{split}
\end{equation}
The four-point amplitude $\mathcal{A}_4$ is determined in momentum space by the soft theorem:
\begin{equation} \label{A4-def}
    \begin{split}
	\mathcal{A}_4(p_1,\hat q, p_3,p_4) 
		&=  \sum_{k = 1,3,4} Q_k \frac{\varepsilon_+ \cdot p_k}{\hat q\cdot p_k} ~ \mathcal{A}_3(p_1, p_3, p_4) 
		=  \sum_{k = 1,3,4} Q_k \frac{\varepsilon_+ \cdot p_k}{\hat q\cdot p_k} ~ g \delta^{(4)} (p_1+p_3+p_4).
    \end{split}
\end{equation}
Parametrizing the momenta as in \eqref{4-point-red}, we find  
\begin{equation} \label{A4-expanded}
    \begin{split} 
	\mathcal{A}_4&( \eps_1 \omega_1 \hat q (0,0), \hat q(z, \bar{z}), \eps_3 \omega_3 \hat q(1,1), \eps_4 m \hat p_4)\\
	& = \frac{1}{\sqrt{2}}\left[ \frac{Q_1}{z} + \frac{Q_3}{z-1} + \frac{Q_4(\bar{z}- \bar{w}_4)}{y_4^2 + |z- w_4|^2}\right]g \delta^{(4)} \left(\eps_1  \omega_1 \hat q(0,0)+ \eps_3 \omega_3 \hat q(1,1)+ \frac{\eps_4m }{2y_4} (y_4^2 n  + \hat q  (w_4, \bar{w}_4) )\right).
    \end{split}
\end{equation}
Notice, the same delta function appears here as in the expression for the three-point function \eqref{3-point-eval} and thus is also solved by \eqref{delta-solution}.  Using \eqref{delta-identity} to evaluate the integrals, \eqref{4-point-red} becomes
\begin{equation} 
    \begin{split}
	&\lim_{\tau, \bar{\tau} \to \infty}  \tau^{\Delta_4} \bar{\tau}^{\Delta_4} 
        \langle \phi_{\Delta_1}(0,0) J(z,\bar{z})
                \phi_{\Delta_3}(1,1) \Phi_{\Delta_4}(\tau,\bar{\tau})\rangle\\
	& \quad  = \frac{g}{4} \left(\frac{m}{2}\right)^{\Delta_1+\Delta_3-4} \int_0^\infty \frac{d \omega_1}{\omega_1} ~ \omega_1^{\Delta_1-\Delta_3+\Delta_4}\left(1 + \omega_1^2 \right)^{-\Delta_4 }  \left[ \frac{Q_1}{z} + \frac{Q_3}{z-1} + \frac{Q_4 ( \bar{z}(1+\omega_1^2)-1)}{(z-1)(\bar{z}-1)+  \omega_1^2  z \bar{z}  }\right] .
    \end{split}
\end{equation} 
Including the prefactors in \eqref{4-point-gen}, making the additional change of variables
\begin{equation}
    \begin{split}
	t = \frac{\omega_1^2}{1+\omega_1^2}, 
    \end{split}
\end{equation}
and recalling the expression for the three-point amplitude \eqref{3-point-final}, we arrive at the following expression for the conformally soft theorem:
\begin{equation} \label{4-point-pre-int}
    \begin{split}
	\langle   \phi_{\Delta_1}&  (z_1, \bar{z}_1)  J(z_2,\bar{z}_2 )
        \phi_{\Delta_3}  (z_3, \bar{z}_3) \Phi_{\Delta_4}   (z_4, \bar{z}_4)\rangle 
	\\	& =  \frac{z_{41}z_{34}}{z_{13} z_{24}^2}
            \frac{ \langle   \phi_{\Delta_1}  (z_1, \bar{z}_1) \phi_{\Delta_3}  (z_3, \bar{z}_3) \Phi_{\Delta_4}   (z_4, \bar{z}_4)\rangle}{B \left(\frac{\Delta_4+\Delta_1-\Delta_3}{2},\frac{\Delta_4-\Delta_1+\Delta_3}{2}\right)}  \int_0^1 \frac{dt}{t(1-t)} ~ t^{\frac{1}{2}(\Delta_4+\Delta_1-\Delta_3)}
            (1-t)^{\frac{1}{2}(\Delta_4-\Delta_1+\Delta_3)}
        \\&  \quad \quad \quad \quad \quad \quad \quad 
            \quad \quad \quad \quad \quad \quad \quad \quad \quad \times  \left[ \frac{Q_1}{z} + \frac{Q_3}{z-1} + \frac{Q_4}{(z-1)(\bar{z}-1)} \frac{  \bar{z}-1+ t}{1- ( 1- \frac{z \bar{z}}{(z-1)(\bar{z}-1)})t  }\right].
    \end{split}
\end{equation}

Before performing the remaining integral, let us pause for a moment and examine the singularity structure in $z_2$.  First, note that the explicit dependence on $\bar{z}$ reveals that in the presence of massive particles, correlators of $J$ are no longer meromorphic functions, despite the fact that $J$ is a weight $(1,0)$ current.  As a result, in $(3,1)$ signature where $z_k$ and $\bar{z}_k$ are related by complex conjugation, one can no longer freely deform contours of integration with respect to $z_2$.  Thus, if we construct charges from line integrals of $J$, then, strictly speaking, standard arguments about contour deformations do not apply in this context. 
  
Nevertheless, if we relax the signature and allow $z_k$ and $\bar{z}_k$ to be independent complex variables, then the correlator is meromorphic in $z_2$ and ${\rm SL} \times {\rm SL}$ symmetry guarantees that the physics of charge conservation is still encoded in the singularity structure in $z_2$.  To see this, consider the following contour integral:
\begin{equation}
    \begin{split}
	\oint_{\mathcal{C}_\infty}\frac{dz}{2 \pi i } \langle J(z, \bar{z}) \mathcal {O}_1(z_1, \bar{z}_1) \cdots  \mathcal{O}_n(z_n, \bar{z}_n)\rangle,
    \end{split}
\end{equation}
where $\mathcal{C}_\infty$ is a closed contour encircling the point at infinity.  We can evaluate this integral in two ways.  First, we can deform the integral to encircle the singularities in the correlator, denoted collectively by $z_j$:
\begin{equation} \label{eval-1}
    \begin{split}
	\oint_{\mathcal{C}_\infty}\frac{dz}{2 \pi i } \langle J(z, \bar{z}) \mathcal {O}_1(z_1, \bar{z}_1) \cdots  \mathcal{O}_n(z_n, \bar{z}_n)\rangle
		&= \sum_{j}\oint_{\mathcal{C}_{z_j}} \frac{dz}{2 \pi i} \langle J(z, \bar{z}) \mathcal {O}_1(z_1, \bar{z}_1) \cdots  \mathcal{O}_n(z_n, \bar{z}_n)\rangle.
    \end{split}
\end{equation}
Note that generically there may be both pointlike and extended singularities, so it is a slight abuse of notation to label them by individual points $z_j$ on the complex plane. Next, we can use the requirement from conformal symmetry that any correlator involving $J$ must fall off as\footnote{A derivation of \eqref{J-conformal} is presented in the main text.  See the discussion leading up to  \eqref{conformal-fall-off}.}
\begin{equation} \label{J-conformal}
    \lim_{z  \to \infty}\langle  J(z, \bar{z}) \mathcal {O}_1(z_1, \bar{z}_1) \cdots  \mathcal{O}_n(z_n, \bar{z}_n)\rangle \sim \frac{1}{z^2}
\end{equation} 	to argue that this integral must vanish:
\begin{equation} \label{eval-2}
    \begin{split}
	\oint_{\mathcal{C}_\infty}\frac{dz}{2 \pi i } \langle J(z, \bar{z}) \mathcal {O}_1(z_1, \bar{z}_1) \cdots  \mathcal{O}_n(z_n, \bar{z}_n)\rangle= 0.
    \end{split}
\end{equation} 
Thus, \eqref{eval-1} and \eqref{eval-2} together imply a constraint on the singularity structure in $z$.

It is instructive to verify this behavior explicitly in our example.  Let us focus on the following contribution  to the integrand in \eqref{4-point-pre-int}:
\begin{equation} \label{mom-z2}
    \begin{split}
         &\frac{z_{41}z_{34}}{z_{13} z_{24}^2}\left[ \frac{Q_1}{z} + \frac{Q_3}{z-1} + \frac{Q_4}{(z-1)(\bar{z}-1)} \frac{  \bar{z}-1+ t}{1- ( 1- \frac{z \bar{z}}{(z-1)(\bar{z}-1)})t  }\right],
    \end{split}
\end{equation}
which contains all the non-trivial $z_2$ dependence.  We  focus on the singularity structure in $z_2$, holding $\bar{z}_2$ fixed.  Note that studying the singularity structure before taking the integral with respect to $t$ effectively amounts to working in momentum space.  Specifically, a similar argument directly applies to $\mathcal{A}_4$ in \eqref{A4-def}. 

The integrand \eqref{mom-z2} contains simple poles in $z_{2}$ at $z_1$ and $z_3$. Near these points, the conformal cross ratio $z$ approaches $0$ and $1$, respectively, and we find
\begin{equation}
    \begin{split}
        \lim_{z_2 \to z_1}\frac{z_{41}z_{34}}{z_{13} z_{24}^2}\left[ \frac{Q_1}{z} + \frac{Q_3}{z-1} + \frac{Q_4}{(z-1)(\bar{z}-1)} \frac{  \bar{z}-1+ t}{1- ( 1- \frac{z \bar{z}}{(z-1)(\bar{z}-1)})t  }\right]
        &\to \frac{Q_1}{z_{21}}, \\
        \lim_{z_2 \to z_3}\frac{z_{41}z_{34}}{z_{13} z_{24}^2}\left[ \frac{Q_1}{z} + \frac{Q_3}{z-1} + \frac{Q_4}{(z-1)(\bar{z}-1)} \frac{  \bar{z}-1+ t}{1- ( 1- \frac{z \bar{z}}{(z-1)(\bar{z}-1)})t  }\right]
        &\to \frac{Q_3}{z_{23}}.
    \end{split}
\end{equation}
Here, it is also helpful to notice that
\begin{equation} \label{eq:standardsing}
\frac{z_{41}z_{34}}{z_{13} z_{24}^2}\left[ \frac{Q_1}{z} + \frac{Q_3}{z-1}\right] = \frac{Q_1}{z_{21}} + \frac{Q_3}{z_{23}} - \frac{Q_1 + Q_3}{z_{24}}.
\end{equation}
In a standard local 2D conformal field theory with charge conservation such that $Q_4 = -(Q_1 +Q_3)$, these terms would constitute the complete singularity structure. In particular, as $z_2$ approaches $z_4$, there would be a simple pole with residue $Q_4$.  However, in \eqref{mom-z2} there is an additional term involving $Q_4$, which also contributes a simple pole in the limit as $z_2$ approaches $z_4$:  
\begin{equation}
    \begin{split}
         \lim_{z_2 \to z_4}\frac{z_{41}z_{34}}{z_{13} z_{24}^2}\left[ \frac{Q_1}{z} + \frac{Q_3}{z-1} + \frac{Q_4}{(z-1)(\bar{z}-1)} \frac{  \bar{z}-1+ t}{1- ( 1- \frac{z \bar{z}}{(z-1)(\bar{z}-1)})t  }\right]
            \to - \frac{Q_1+ Q_3 + Q_4}{z_{24}} = 0.
    \end{split}
\end{equation}
Thus, unlike in a standard 2D conformal field theory, the full expression is regular at $z_4$. Notice that \eqref{eval-2} implies that \eqref{mom-z2} must contain another singularity in $z_2$, since the residues considered so far do not add to zero. Indeed, the final singularity is the solution to
\begin{equation}
    1- \left[ 1- \frac{z \bar{z}}{(z-1)(\bar{z}-1)}\right]t = 0.
\end{equation}
Denoting the solution by $z_*$, where 
\begin{equation} \label{mom-pole}
    z_*  =  \frac{t \bar{z} z_1 z_{34} - (1-t) (\bar{z}-1)z_3 z_{41}}{t \bar{z}z_{34}-(1-t)(\bar{z}-1)z_{41}},
\end{equation}
we find
\begin{equation} \label{mom-last-pole}
    \lim_{z_2 \to z_*}\frac{z_{41}z_{34}}{z_{13} z_{24}^2}\left[ \frac{Q_1}{z} + \frac{Q_3}{z-1} + \frac{Q_4}{(z-1)(\bar{z}-1)} \frac{  \bar{z}-1+ t}{1- ( 1- \frac{z \bar{z}}{(z-1)(\bar{z}-1)})t  }\right]  \to \frac{Q_4}{z_{2*}}.
\end{equation}
The explicit dependence of $z_*$ on $t$ in \eqref{mom-pole} reflects the fact that in complexified momentum space, a null momentum $\hat q$ (satisfying $\hat q^2 = 0$) can have vanishing inner product with a timelike momentum $p_k$ (satisfying $p_k^2 <0$). That is, the simple pole at $z_*$ is the pole that arises when the inner product between $\hat q(z_2, \bar{z}_2)$ and $p_4 = m \hat p_4$ vanishes. 

Consistency requires that the singularity structure of the celestial amplitude in $z_2$ also conspires to respect \eqref{eval-2} after performing the final integral in $t$. To investigate this behavior, we use the integral definition of the hypergeometric function
\begin{equation} \label{hypergeo-int}
    \begin{split}
	B(b,c-b) {}_2F_1(a,b,c;x) = \int_0^1 dt ~ t^{b-1} (1-t)^{c-b-1}(1-xt)^{-a}, 
    \end{split}
\end{equation}
and the identity
\begin{equation} 
    {}_2 F_1(a, b, c;x) - 
        {}_2 F_1(a-1, b, c;x) - \frac{b}{c} x ~{}_2 F_1(a, b+1,c+1;x) =0 
\end{equation}
to obtain the following closed form expression for the conformally soft theorem:
\begin{equation} \label{4-point-post-int-mina2}
    \begin{split}
	\langle   \phi_{\Delta_1}&  (z_1, \bar{z}_1)  J(z_2,\bar{z}_2 )
        \phi_{\Delta_3}  (z_3, \bar{z}_3) \Phi_{\Delta_4}   (z_4, \bar{z}_4)\rangle 
	\\	&=  \Bigg[ \frac{Q_1}{z_{21}} + \frac{Q_3}{z_{23}} + \frac{Q_4}{z_{24}}\Bigg(1 + \frac{z_{34}}{z_{23}}  \frac{1}{z{-}1{+}\bar{z}} \left[z{-}1 {+} \bar{z} \,{}_2F_1 \Big(1,\frac{\Delta_1{-}\Delta_3{+}\Delta_4}{2},\Delta_4;1{-} \frac{z}{z{-}1}\frac{\bar{z}}{\bar{z}{-}1} \Big) \right] \Bigg)  \Bigg]
        \\&  \quad \quad \quad \quad  \quad  \quad \quad \quad \quad \quad \quad \quad \quad \quad \quad \quad \quad \quad \quad \quad \quad \quad \quad \times \langle   \phi_{\Delta_1}  (z_1, \bar{z}_1) \phi_{\Delta_3}  (z_3, \bar{z}_3) \Phi_{\Delta_4}   (z_4, \bar{z}_4)\rangle.
	\end{split}
\end{equation}
The hypergeometric integral \eqref{hypergeo-int} converges provided ${\rm Re}(c)> {\rm Re}(b) >0$ so here we assume that the conformal weights respect
\begin{equation}
    {\rm Re}(\Delta_4)> {\rm Re}\left(\frac{\Delta_1-\Delta_3+\Delta_4}{2}\right) >0.
\end{equation}
Notice that the spectrum of conformal weights presented in \cite{Pasterski:2017kqt} satisfies this condition. In the conformal primary basis, \eqref{4-point-post-int-mina2} retains the simple poles in $z_2$ at $z_1$ and $z_3$ with residues proportional to $Q_1$ and $Q_3$, respectively:
\begin{equation}
    \begin{split}
        \lim_{z_{2} \to z_1} z_{21}
            \frac{\langle   \phi_{\Delta_1}   (z_1, \bar{z}_1)  J(z_2,\bar{z}_2 )
        \phi_{\Delta_3}  (z_3, \bar{z}_3) \Phi_{\Delta_4}   (z_4, \bar{z}_4)\rangle }{\langle   \phi_{\Delta_1}  (z_1, \bar{z}_1) \phi_{\Delta_3}  (z_3, \bar{z}_3) \Phi_{\Delta_4}   (z_4, \bar{z}_4)\rangle}
             = Q_1, \\
             \lim_{z_{2} \to z_3} z_{23}
        \frac{\langle   \phi_{\Delta_1}   (z_1, \bar{z}_1)  J(z_2,\bar{z}_2 )
        \phi_{\Delta_3}  (z_3, \bar{z}_3) \Phi_{\Delta_4}   (z_4, \bar{z}_4)\rangle }{\langle   \phi_{\Delta_1}  (z_1, \bar{z}_1) \phi_{\Delta_3}  (z_3, \bar{z}_3) \Phi_{\Delta_4}   (z_4, \bar{z}_4)\rangle}
             = Q_3.
    \end{split}
\end{equation}
For the second limit, we use $z_{23} \to 0$, which corresponds to $z\to 1$ and sends the argument of the hypergeometric function to infinity. Note that the hypergeometric function has branch points at the arguments $x=1$ and $x = \infty$, and a branch cut extends between them. Generic paths $z_{23}\to 0$ avoid this branch cut and in this limit
\begin{equation}
    \begin{split}
        \lim_{|x| \to \infty} &B\left(\frac{\Delta_1-\Delta_3+\Delta_4}{2},\frac{\Delta_3-\Delta_1+\Delta_4}{2}\right)
        {}_2F_1 \left(1,\frac{\Delta_1-\Delta_3+\Delta_4}{2},\Delta_4;x\right)\\
            &  \quad \quad \quad = \lim_{|x| \to \infty}  \int_0^1 \frac{dt}{t(1-t)} ~ t^{\frac{1}{2}(\Delta_4+\Delta_1-\Delta_3)}
            (1-t)^{\frac{1}{2}(\Delta_4-\Delta_1+\Delta_3)} \frac{1}{1-xt}  
             = 0. 
    \end{split} 
\end{equation}
Similarly, for generic paths $z_{21}\to 0$ that avoid the branch cut, the hypergeometric function is regular. Also, as before, there is no simple pole in \eqref{4-point-post-int-mina2} at $z_4$. To see this, recall that $z_2 \to z_4$ implies $z \to \infty$, so we find
\begin{equation}
    \begin{split}
       \lim_{z_2 \to z_4}& \frac{Q_4}{z_{24}}\Bigg(1 + \frac{z_{34}}{z_{23}}  \frac{1}{z{-}1{+}\bar{z}} \left[z{-}1 {+} \bar{z} \,{}_2F_1 \Big(1,\frac{\Delta_1{-}\Delta_3{+}\Delta_4}{2},\Delta_4;1{-} \frac{z}{z{-}1}\frac{\bar{z}}{\bar{z}{-}1} \Big) \right]\Bigg)\\
        &=\lim_{z_2 \to z_4} \frac{Q_4}{z_{24}}\left(1 + \frac{z_{34}}{z_{23}} \right)
         = \frac{Q_4}{z_{43}},
    \end{split}
\end{equation}
which is manifestly regular. Finally, as in momentum space, there naively appears to be another pole in $z_2$, this time where
\begin{equation}
    z-1+ \bar{z} \to 0,
\end{equation}
or equivalently, 
\begin{equation}
     z_2 \to \frac{z_1z_{34}+ z_4z_{13} (1- \bar{z})}{z_{34}+z_{13}(1- \bar{z})} .
\end{equation}
However, in this limit, the hypergeometric function approaches unity
\begin{equation}
    \lim_{z \to 1- \bar{z}} 
         {}_2F_1 \Big(1,\frac{\Delta_1-\Delta_3+\Delta_4}{2},\Delta_4;1- \frac{z}{z-1}\frac{\bar{z}}{\bar{z}-1} \Big)
          =  {}_2F_1 \Big(1,\frac{\Delta_1-\Delta_3+\Delta_4}{2},\Delta_4;0\Big) = 1,
\end{equation}
so there is no simple pole in $z_2$ at this point. 

Specifying \eqref{eval-2} to the four-point amplitude \eqref{4-point-post-int-mina2} and deforming the contour, we again reach the conclusion that there must be some other singularity in $z_2$.  We will now demonstrate that the contribution from the third simple pole in momentum space \eqref{mom-last-pole} is replaced by a contribution from the discontinuity along the branch cut of the hypergeometric function, so that 
\begin{equation} \label{contour-expanded-mina2}
    \begin{split}
        0&= \oint_{\mathcal{C}_\infty} \frac{dz_2}{2 \pi i}
             \langle  \phi_{\Delta_1} (z_1)  J(z_2 )
        \phi_{\Delta_3}  (z_3) \Phi_{\Delta_4}   (z_4) \rangle\\
         &= {\rm Res}_{z_1} \left[\langle  \phi_{\Delta_1}   (z_1)  J(z_2 )
        \phi_{\Delta_3}  (z_3) \Phi_{\Delta_4}   (z_4)  \rangle \right] 
        +{\rm Res}_{z_3} \left[\langle  \phi_{\Delta_1}   (z_1)  J(z_2 )
        \phi_{\Delta_3}  (z_3) \Phi_{\Delta_4}   (z_4) \rangle \right]\\& 
        \quad \quad + \int_{\mathcal{C}} \frac{dz_2}{2 \pi i }
        ~ {\rm disc} \left[ \langle \phi_{\Delta_1}   (z_1)  J(z_2 )
        \phi_{\Delta_3}  (z_3) \Phi_{\Delta_4}   (z_4)  \rangle\right]\\
        & = \left(Q_1 + Q_3\right)\langle  \phi_{\Delta_1}   (z_1) 
        \phi_{\Delta_3}  (z_3) \Phi_{\Delta_4}   (z_4)  \rangle + \int_{\mathcal{C}} \frac{dz_2}{2 \pi i }
        ~ {\rm disc} \left[ \langle \phi_{\Delta_1}   (z_1)  J(z_2 )
        \phi_{\Delta_3}  (z_3) \Phi_{\Delta_4}   (z_4) \rangle\right],
    \end{split}
\end{equation}
where $\mathcal{C}$ is a contour along the branch cut in $z_2$ between $z_1$ and $z_3$. Choosing the contour around infinity to have clockwise orientation as usual, it can be deformed to contours around each simple pole and the branch cut with counterclockwise orientation.

To determine the contribution from the cut, first note that we can take the branch cut in the hypergeometric function to extend along the positive real $x$-axis from $x=1$ to $x = \infty$ and the discontinuity to be defined as the negative imaginary contour subtracted from the positive imaginary contour along this axis. The contribution to contour integral is\footnote{The orientation of the $x$ contour can be deduced from the counterclockwise contour orientation in $z_2$ by  mapping $z_1 \to 0, z_3 \to 1, z_4 \to \infty$.}
\begin{equation} \label{branch-contour}
    \begin{split}  
        \int_{\mathcal{C}} \frac{dz_2}{2 \pi i}~{\rm disc}&\left[ \frac{z_{34}}{z_{23} z_{24}}\frac{ Q_4\bar{z} }{z-1+\bar{z}}
              {}_2F_1 \Big(1,\frac{\Delta_1-\Delta_3+\Delta_4}{2},\Delta_4;1- \frac{z}{z-1}\frac{\bar{z}}{\bar{z}-1} \Big) \right]\\
        &  \quad \quad \quad \quad \quad \quad \quad \quad \quad =  \frac{Q_4}{2 \pi i} 
        \int_1^\infty \frac{dx}{x}~ 
        {\rm disc}\left[
              {}_2F_1 \Big(1,\frac{\Delta_1-\Delta_3+\Delta_4}{2},\Delta_4;x \Big) \right], 
    \end{split}
\end{equation}
where we simplified using the change of variables
\begin{equation}
    x = 1- \frac{z}{z-1}\frac{\bar{z}}{\bar{z}-1}, 
    \quad \quad dx =  \frac{z_{13}z_{34}}{z_{41}z_{23}^2} 
   \frac{\bar{z}}{\bar{z}-1} dz_2.
\end{equation} 
The value of the discontinuity can be determined from the integral form \eqref{hypergeo-int}:
\begin{equation}
    \begin{split}
        {\rm disc} \left[{}_2F_1(1, b, c; x)\right] 
            & = \frac{1}{B(b,c-b)}
                \int_0^1 \frac{du}{u(1-u)}~ u^b (1-u)^{c-b}
                  \lim_{\eps \to 0} \left[ \frac{1}{1- u x- i \eps}- \frac{1}{1- u x+ i \eps}\right]\\
            & =\frac{1}{B(b,c-b)}
                \int_0^1 \frac{du}{u(1-u)}~ u^b (1-u)^{c-b}
                    2 \pi i \delta(1-ux)\\
            & =\frac{2 \pi i}{B(b,c-b)} \Theta(x-1)~ x^{-c+1} (x-1)^{c-b-1}.
    \end{split}
\end{equation}
Substituting back into \eqref{branch-contour}, we find 
\begin{equation}  
    \begin{split} 
        \int_{\mathcal{C}} \frac{dz_2}{2 \pi i}~{\rm disc}&\left[ \frac{z_{34}}{z_{23} z_{24}}\frac{ Q_4\bar{z} }{z-1+\bar{z}}
              {}_2F_1 \Big(1,\frac{\Delta_1-\Delta_3+\Delta_4}{2},\Delta_4;1- \frac{z}{z-1}\frac{\bar{z}}{\bar{z}-1} \Big) \right] = Q_4,
    \end{split}
\end{equation}
which is precisely the contribution needed for the consistency of \eqref{contour-expanded-mina2}. 

The main lesson to take away from this appendix is that celestial operators representing massive particles can admit non-local behavior.  In particular, in theories with charged massive particles, it appears that charge conservation in celestial holography is still consistent with contour integrals of soft photon currents, but massive particles may produce branch cut singularities instead of the simple poles that are familiar from the context of a local conformal field theory. 

\section{${\rm SL}(2, \mathbb{C})$ action on bulk and boundary points}\label{app_b}
	 
In this appendix, we justify the statement in the introduction that the set of ${\rm SL}(2, \mathbb{C})$ transformations that preserve a point on the celestial sphere preserve no proper subset of points in the bulk of ${\rm AdS}_3$.  For simplicity and without loss of generality, consider the conformal transformations that preserve the origin on the celestial plane $(z, \bar{z}) = (0,0)$.  These take the form of \eqref{sl2c} with
\begin{equation}
    \begin{split}
	\left( \begin{matrix}a & b\\ c&d \end{matrix} \right) = \left( \begin{matrix}1/d & 0\\ c&d \end{matrix} \right).
    \end{split}
\end{equation}
Selecting an arbitrary bulk point $(y_k, w_k, \bar{w}_k)$, notice that the transformation with 
\begin{equation}
    \begin{split}
	\left(c,d \right) = \sqrt{\frac{y_k}{y_k^2 + w_k \bar{w}_k}} \left(\bar{w}_k/y_k , 1\right),
    \end{split}
\end{equation}
maps the bulk point $(1,0,0)$ to this point $(y_k,w_k, \bar{w}_k)$. Likewise, the inverse transformation maps the arbitrarily point  $(y_k,w_k, \bar{w}_k)$ to $(1,0,0)$.  Therefore by suitably combining such transformations, we can find an ${\rm SL}(2, \mathbb{C})$ transformation that preserves the boundary point $(z, \bar{z}) = (0,0)$, but maps between any pair of bulk points.

\section{Derivation of massive generating function} \label{app:massivegen} 

In this appendix, we include more details about the derivation of \eqref{eq:massivegenfun}. Starting from \eqref{eq:massivemidstep} and using \eqref{eq:Fpm} we find:
\begin{equation}
  \begin{aligned}
    \mathbf{G}_p(\bar{z}) &=  - \frac{\mathcal{N}_{2p-4}\mathcal{C}_p}{2 (2p-2)! m_k^{4p-6}} \frac{ (-1)^{2p-3}}{(4p-6)! m_k^{4p-6}} \\
    &\qquad \times \sum_{n=0}^{4p-6} \left(-\partial_{z}\right)^{2p-2-n} (\hat q \cdot   p_k)^{2p-2} \partial_z^n \left[\sum_{\ell = 0}^{4p-6} \left[(-\partial_{\bar{z}})^{4p-6-\ell} (p_k \cdot \hat q )^{4p-6}\right]~ \partial_{\bar{z}}^{\ell}\left( F_+ \cdot \mathcal{L}_k \right)^{2p-3} \right] . 
  \end{aligned}
\end{equation}
Now consider the sum
\begin{equation}
  S = \sum_{n=0}^{4p-6}\left(- \partial_z\right)^{2p-2 -n}\left(\hat{q}\cdot p_k\right)^{2p-2} \partial_z^n \left[\sum_{\ell=0}^{4p-6}\left(-\partial_{\bar{z}}\right)^{4p-6-\ell}\left(\hat{q}\cdot p_k\right)^{4p-6}\partial_{\bar{z}}^{\ell}\left( F_+ \cdot \mathcal{L}_k\right)^{2p-3} \right]. 
\end{equation}
Using completeness \eqref{completeness1} and the fact that
\begin{equation}
  \partial_{z}^n\left(\partial_{\bar{z}}\hat{q}\cdot p_k\right)^{m} = \frac{m!}{(m-n)!} \left(\partial_{\bar{z}}\hat{q}\cdot p_k\right)^{m-n}\left(n \cdot p_k\right)^{n}, 
\end{equation}
which is formally true for both positive and negative $n$, we can rewrite this as
\begin{equation} \label{eq:giantsum}
  \begin{aligned}
    S &=  \sum_{\ell=0}^{4p-6}(m_k^2)^{4p-6}(-1)^{\ell} \frac{\left(\partial_z \hat{q}\cdot p_k\right)^{\ell - 2p + 4}}{(n \cdot p_k)^{\ell}} \partial_{\bar{z}}^{\ell}  \left(F_+ \cdot \mathcal{L}_k\right)^{2p-3} \\
    &\qquad \times \left[\sum_{n=0}^{4p-6}\sum_{m=0}^{2p-2} \sum_{j=0}^{4p-6} {2p-2 \choose m} {4p-6 \choose j} \frac{(-1)^{6p-8-n} m! j! j! \left(\frac{\partial_{\bar{z}}\hat{q}\cdot p_k \partial_{z}\hat{q}\cdot p_k}{m_k^2}\right)^{j+m-2p+2}}{(m-(2p-2-n))!(j -( 4p-6 -\ell))!(j-n)!} \right].
  \end{aligned}
\end{equation}
To simplify the sum in brackets, consider the sum
\begin{equation}
  S' = \sum_{n=0}^{4p-6}\sum_{m=0}^{2p-2} \sum_{j=0}^{4p-6} (-1)^{6p-8-n}{2p-2 \choose m} {4p-6 \choose j} \frac{m! j! j! A^{j+m-2p+2}}{(m-(2p-2-n))!(j -( 4p-6 -\ell))!(j-n)!},
\end{equation}
where $A$ does not depend on the indices in the sum. We re-index with $k = m+j$ and evaluate the sum in $n$ first: 
\begin{equation} \label{eq:Sprime}
  \begin{aligned}
    S' 
    &= \sum_{j=0}^{4p-6} \sum_{k=j}^{2p-2+j}   \frac{(-1)^{2p}}{(k-2p+2)} \frac{(2p-2)!(4p-6)!A^{k-2p+2}}{(k-j-2p+1)!(2p-2-k+j)!(4p-6-j)!(j-4p+6+\ell)!}. 
  \end{aligned}
\end{equation}
Notice that if $k-j < 2p+1$, the first factorial in the denominator diverges, while if $k-j > 2p -2$, the second factorial in the denominator diverges. This will only be cancelled by the divergence in $\frac{1}{(k-2p+2)}$ if $k=2p-2$, which also enforces $j \leq 2p-2$ since $k = m+j$. Therefore, the sum collapses and we find, including normalization, that
\begin{equation}
  \begin{aligned} \mathbf{G}_p(\bar{z}) =  \frac{\mathcal{N}_{2p-4}\mathcal{C}_p}{2  }\frac{1}{\left(-n \cdot p_k\right)^{2p-4}}  \sum_{\ell=0}^{4p-6} \frac{1}{\ell!} \left(\sum_{j=2p-4}^{4p-6} (-1)^j {\ell \choose j} \right)\left(- \frac{\partial_z \hat{q}\cdot p_k}{n\cdot p_k} \right)^{\ell - 2p + 4} \partial_{\bar{z}}^{\ell}  \left(F_+ \cdot \mathcal{L}_k\right)^{2p-3} ,  
  \end{aligned}
\end{equation}
which is manifestly zero when $\ell < 2p-4$. Note that for $p > 2$ and $\ell \leq 4p-6$ we have the simplification
\begin{equation}
  \sum_{j=2p-4}^{4p-6} (-1)^j {\ell \choose j} = (-1)^{2p}\frac{(2p-4)}{\ell}{\ell \choose 2p -4}, \ \ \ p  > 2 \ {\rm and } \  \ell \leq 4p-6,
\end{equation}
which is used to obtain the simplified form \eqref{eq:massivegenspec}. 

\section{Details of massive momentum-space action} \label{app:w52}

In this appendix, we provide details of the proof of the base case \eqref{eq:w52proof}.  First, note that if we demonstrate that 
\begin{equation} \label{eq:w52full}
    \begin{split}
        \left[ \delta^{\frac{5}{2}}_{m},  \delta^q_{n}\right] |p_k \rangle 
        &  =\left(m(q-1) - \frac{3}{2}n\right) \delta^{q+\frac{1}{2}}_{m+n}|p_k \rangle
    \end{split}
\end{equation}
holds for all modes $-\frac{3}{2} \leq m \leq \frac{3}{2}$ and a single mode $n \in \left[1-q,q-1\right]$, then using the action of $\delta^{2}_k$ \eqref{eq:delta2comm}, it can be shown that it also holds for all $n \in \left[1-q,q-1\right]$. This statement is a consequence of $\rm{SL}(2,\mathbb{C})$ symmetry and the proof is exactly as in Appendix C.2 of \cite{Himwich:2021dau}. Then $\eqref{eq:w52proof}$ will follow as a special case of \eqref{eq:w52full} with $m=-\frac{3}{2}$.  We will therefore be done if we show that 
\begin{equation} \label{eq:w52proofstep}
    \begin{split}
        \left[ \delta^{\frac{5}{2}}_{m},  \delta^q_{1-q}\right] |p_k \rangle 
        &  =(q-1)\left(m + \frac{3}{2} \right) \delta^{q+\frac{1}{2}}_{1+m-q}|p_k \rangle
    \end{split}
\end{equation}
for $-\frac{3}{2} \leq m \leq \frac{3}{2}$.  To do so, we use the explicit form of $\delta^{q+\frac{1}{2}}_{1-q+m}$ for $-\frac{3}{2} \leq m \leq \frac{3}{2}$, which can be derived from the integral form \eqref{massive-tf4} or equivalently from the generating function \eqref{eq:massivegenfun}. The modes relevant to the commutator \eqref{eq:w52proofstep} have action 
\begin{equation}
\begin{aligned}
\delta^{q}_{1-q} |p_k\rangle &= \frac{1}{2}\left(\frac{2 y_k}{\epsilon_k m_k}\right)^{2q-4} (2q-2)\partial_{\bar{w}_k}^{2q-3}|p_k \rangle, \\
\delta^{q}_{2-q} |p_k\rangle &= \frac{1}{2} \left[(2q-2) \bar{w}_k \partial_{\bar{w}_k} + (2q-3)y_k \partial_{y_k}\right]\left(\frac{2 y_k}{\epsilon_k m_k}\right)^{2q-4} \partial_{\bar{w}_k}^{2q-4} |p_k \rangle, \\
\delta^{q}_{3-q} |p_k\rangle &= \frac{1}{2} \left[(2q-2) \bar{w}_k^2 \partial_{\bar{w}_k}^2 + 2(2q-3)y_k \partial_{y_k}\bar{w}_k \partial_{\bar{w}_k} + (2q-4)y_k\partial_{y_k}(y_k \partial_{y_k} - 1) - 2y_k^2\partial_{w_k}\partial_{\bar{w}_k}\right] \\
&\qquad \qquad \qquad \qquad \qquad \qquad \qquad \qquad \qquad \qquad \qquad \qquad \qquad  \times \left(\frac{2 y_k}{\epsilon_k m_k}\right)^{2q-4} \partial_{\bar{w}_k}^{2q-5} |p_k \rangle, \\
\delta^{q}_{4-q} |p_k\rangle &= \frac{1}{2} \Big[(2q-2) \bar{w}_k^3 \partial_{\bar{w}_k}^3 + 3(2q-3)y_k \partial_{y_k}\bar{w}_k^2 \partial_{\bar{w}_k}^2 + 3(2q-4)y_k\partial_{y_k}(y_k \partial_{y_k} - 1)\bar{w}_k \partial_{\bar{w}_k} \\
&\qquad + (2q-5)y_k\partial_{y_k}(y_k \partial_{y_k} - 1)(y_k \partial_{y_k} - 2) - 6y_k^2\bar{w_k}\partial_{\bar{w}_k}^2\partial_{w_k} - 6y_k^3\partial_{y_k}\partial_{w_k}\partial_{\bar{w}_k}\Big] \\
&\qquad \qquad \qquad \qquad \qquad \qquad \qquad \qquad \qquad \qquad \qquad \qquad \qquad  \times \left(\frac{2 y_k}{\epsilon_k m_k}\right)^{2q-4} \partial_{\bar{w}_k}^{2q-6} |p_k \rangle.  \\
\end{aligned}
\end{equation}
Using this form, one can verify \eqref{eq:w52proofstep} for $-\frac{3}{2} \leq m \leq \frac{3}{2}$ by brute force, completing the proof.        	 			
\end{appendix}	
	
\bibliography{massive}

\providecommand{\href}[2]{#2}\begingroup\raggedright\begin{thebibliography}{10}

\bibitem{Maldacena:1997re}
J.~M. Maldacena, ``{The Large N limit of superconformal field theories and
  supergravity},'' \href{http://dx.doi.org/10.1023/A:1026654312961}{{\em Adv.
  Theor. Math. Phys.} {\bfseries 2} (1998) 231--252},
  \href{http://arxiv.org/abs/hep-th/9711200}{{\ttfamily arXiv:hep-th/9711200}}.

\bibitem{Witten:1998qj}
E.~Witten, ``{Anti-de Sitter space and holography},''
  \href{http://dx.doi.org/10.4310/ATMP.1998.v2.n2.a2}{{\em Adv. Theor. Math.
  Phys.} {\bfseries 2} (1998) 253--291},
  \href{http://arxiv.org/abs/hep-th/9802150}{{\ttfamily arXiv:hep-th/9802150}}.

\bibitem{Heemskerk:2009pn}
I.~Heemskerk, J.~Penedones, J.~Polchinski, and J.~Sully, ``{Holography from
  Conformal Field Theory},''
  \href{http://dx.doi.org/10.1088/1126-6708/2009/10/079}{{\em JHEP} {\bfseries
  10} (2009) 079}, \href{http://arxiv.org/abs/0907.0151}{{\ttfamily
  arXiv:0907.0151 [hep-th]}}.

\bibitem{Maldacena:2015iua}
J.~Maldacena, D.~Simmons-Duffin, and A.~Zhiboedov, ``{Looking for a bulk
  point},'' \href{http://dx.doi.org/10.1007/JHEP01(2017)013}{{\em JHEP}
  {\bfseries 01} (2017) 013}, \href{http://arxiv.org/abs/1509.03612}{{\ttfamily
  arXiv:1509.03612 [hep-th]}}.

\bibitem{Caron-Huot:2021enk}
S.~Caron-Huot, D.~Mazac, L.~Rastelli, and D.~Simmons-Duffin, ``{AdS bulk
  locality from sharp CFT bounds},''
  \href{http://dx.doi.org/10.1007/JHEP11(2021)164}{{\em JHEP} {\bfseries 11}
  (2021) 164}, \href{http://arxiv.org/abs/2106.10274}{{\ttfamily
  arXiv:2106.10274 [hep-th]}}.

\bibitem{Strominger:2001pn}
A.~Strominger, ``{The dS / CFT correspondence},''
  \href{http://dx.doi.org/10.1088/1126-6708/2001/10/034}{{\em JHEP} {\bfseries
  10} (2001) 034}, \href{http://arxiv.org/abs/hep-th/0106113}{{\ttfamily
  arXiv:hep-th/0106113}}.

\bibitem{deBoer:2003vf}
J.~de~Boer and S.~N. Solodukhin, ``{A Holographic reduction of Minkowski
  space-time},'' \href{http://dx.doi.org/10.1016/S0550-3213(03)00494-2}{{\em
  Nucl. Phys. B} {\bfseries 665} (2003) 545--593},
  \href{http://arxiv.org/abs/hep-th/0303006}{{\ttfamily arXiv:hep-th/0303006}}.

\bibitem{Guica:2008mu}
M.~Guica, T.~Hartman, W.~Song, and A.~Strominger, ``{The Kerr/CFT
  Correspondence},'' \href{http://dx.doi.org/10.1103/PhysRevD.80.124008}{{\em
  Phys. Rev. D} {\bfseries 80} (2009) 124008},
  \href{http://arxiv.org/abs/0809.4266}{{\ttfamily arXiv:0809.4266 [hep-th]}}.

\bibitem{Costello:2018zrm}
K.~Costello and D.~Gaiotto, ``{Twisted Holography},''
  \href{http://arxiv.org/abs/1812.09257}{{\ttfamily arXiv:1812.09257
  [hep-th]}}.

\bibitem{Pasterski:2021rjz}
S.~Pasterski, ``{Lectures on celestial amplitudes},''
  \href{http://dx.doi.org/10.1140/epjc/s10052-021-09846-7}{{\em Eur. Phys. J.
  C} {\bfseries 81} no.~12, (2021) 1062},
  \href{http://arxiv.org/abs/2108.04801}{{\ttfamily arXiv:2108.04801
  [hep-th]}}.

\bibitem{Raclariu:2021zjz}
A.-M. Raclariu, ``{Lectures on Celestial Holography},''
  \href{http://arxiv.org/abs/2107.02075}{{\ttfamily arXiv:2107.02075
  [hep-th]}}.

\bibitem{Pasterski:2021raf}
S.~Pasterski, M.~Pate, and A.-M. Raclariu, ``{Celestial Holography},'' in {\em
  {2022 Snowmass Summer Study}}.
\newblock 11, 2021.
\newblock \href{http://arxiv.org/abs/2111.11392}{{\ttfamily arXiv:2111.11392
  [hep-th]}}.

\bibitem{Kapec:2016jld}
D.~Kapec, P.~Mitra, A.-M. Raclariu, and A.~Strominger, ``{2D Stress Tensor for
  4D Gravity},'' \href{http://dx.doi.org/10.1103/PhysRevLett.119.121601}{{\em
  Phys. Rev. Lett.} {\bfseries 119} no.~12, (2017) 121601},
  \href{http://arxiv.org/abs/1609.00282}{{\ttfamily arXiv:1609.00282
  [hep-th]}}.

\bibitem{Cheung:2016iub}
C.~Cheung, A.~de~la Fuente, and R.~Sundrum, ``{4D scattering amplitudes and
  asymptotic symmetries from 2D CFT},''
  \href{http://dx.doi.org/10.1007/JHEP01(2017)112}{{\em JHEP} {\bfseries 01}
  (2017) 112}, \href{http://arxiv.org/abs/1609.00732}{{\ttfamily
  arXiv:1609.00732 [hep-th]}}.

\bibitem{Pasterski:2016qvg}
S.~Pasterski, S.-H. Shao, and A.~Strominger, ``{Flat Space Amplitudes and
  Conformal Symmetry of the Celestial Sphere},''
  \href{http://dx.doi.org/10.1103/PhysRevD.96.065026}{{\em Phys. Rev. D}
  {\bfseries 96} no.~6, (2017) 065026},
  \href{http://arxiv.org/abs/1701.00049}{{\ttfamily arXiv:1701.00049
  [hep-th]}}.

\bibitem{Pasterski:2017kqt}
S.~Pasterski and S.-H. Shao, ``{Conformal basis for flat space amplitudes},''
  \href{http://dx.doi.org/10.1103/PhysRevD.96.065022}{{\em Phys. Rev. D}
  {\bfseries 96} no.~6, (2017) 065022},
  \href{http://arxiv.org/abs/1705.01027}{{\ttfamily arXiv:1705.01027
  [hep-th]}}.

\bibitem{Campiglia:2015qka}
M.~Campiglia and A.~Laddha, ``{Asymptotic symmetries of QED and
  Weinberg\textquoteright{}s soft photon theorem},''
  \href{http://dx.doi.org/10.1007/JHEP07(2015)115}{{\em JHEP} {\bfseries 07}
  (2015) 115}, \href{http://arxiv.org/abs/1505.05346}{{\ttfamily
  arXiv:1505.05346 [hep-th]}}.

\bibitem{Campiglia:2015kxa}
M.~Campiglia and A.~Laddha, ``{Asymptotic symmetries of gravity and soft
  theorems for massive particles},''
  \href{http://dx.doi.org/10.1007/JHEP12(2015)094}{{\em JHEP} {\bfseries 12}
  (2015) 094}, \href{http://arxiv.org/abs/1509.01406}{{\ttfamily
  arXiv:1509.01406 [hep-th]}}.

\bibitem{Freidel:2022skz}
L.~Freidel, D.~Pranzetti, and A.-M. Raclariu, ``{A discrete basis for celestial
  holography},'' \href{http://arxiv.org/abs/2212.12469}{{\ttfamily
  arXiv:2212.12469 [hep-th]}}.

\bibitem{Cotler:2023qwh}
J.~Cotler, N.~Miller, and A.~Strominger, ``{An integer basis for celestial
  amplitudes},'' \href{http://dx.doi.org/10.1007/JHEP08(2023)192}{{\em JHEP}
  {\bfseries 08} (2023) 192}, \href{http://arxiv.org/abs/2302.04905}{{\ttfamily
  arXiv:2302.04905 [hep-th]}}.

\bibitem{Casali:2022fro}
E.~Casali, W.~Melton, and A.~Strominger, ``{Celestial amplitudes as AdS-Witten
  diagrams},'' \href{http://dx.doi.org/10.1007/JHEP11(2022)140}{{\em JHEP}
  {\bfseries 11} (2022) 140}, \href{http://arxiv.org/abs/2204.10249}{{\ttfamily
  arXiv:2204.10249 [hep-th]}}.

\bibitem{Banerjee:2023rni}
S.~Banerjee, R.~Mandal, A.~Manu, and P.~Paul, ``{MHV gluon scattering in the
  massive scalar background and celestial OPE},''
  \href{http://dx.doi.org/10.1007/JHEP10(2023)007}{{\em JHEP} {\bfseries 10}
  (2023) 007}, \href{http://arxiv.org/abs/2302.10245}{{\ttfamily
  arXiv:2302.10245 [hep-th]}}.

\bibitem{Taylor:2023bzj}
T.~R. Taylor and B.~Zhu, ``{Celestial Supersymmetry},''
  \href{http://dx.doi.org/10.1007/JHEP06(2023)210}{{\em JHEP} {\bfseries 06}
  (2023) 210}, \href{http://arxiv.org/abs/2302.12830}{{\ttfamily
  arXiv:2302.12830 [hep-th]}}.

\bibitem{Ball:2023ukj}
A.~Ball, S.~De, A.~Yelleshpur~Srikant, and A.~Volovich, ``{Scalar-Graviton
  Amplitudes and Celestial Holography},''
  \href{http://arxiv.org/abs/2310.00520}{{\ttfamily arXiv:2310.00520
  [hep-th]}}.

\bibitem{Cardona:2017keg}
C.~Cardona and Y.-t. Huang, ``{S-matrix singularities and CFT correlation
  functions},'' \href{http://dx.doi.org/10.1007/JHEP08(2017)133}{{\em JHEP}
  {\bfseries 08} (2017) 133}, \href{http://arxiv.org/abs/1702.03283}{{\ttfamily
  arXiv:1702.03283 [hep-th]}}.

\bibitem{Lam:2017ofc}
H.~T. Lam and S.-H. Shao, ``{Conformal Basis, Optical Theorem, and the Bulk
  Point Singularity},''
  \href{http://dx.doi.org/10.1103/PhysRevD.98.025020}{{\em Phys. Rev. D}
  {\bfseries 98} no.~2, (2018) 025020},
  \href{http://arxiv.org/abs/1711.06138}{{\ttfamily arXiv:1711.06138
  [hep-th]}}.

\bibitem{Nandan:2019jas}
D.~Nandan, A.~Schreiber, A.~Volovich, and M.~Zlotnikov, ``{Celestial
  Amplitudes: Conformal Partial Waves and Soft Limits},''
  \href{http://dx.doi.org/10.1007/JHEP10(2019)018}{{\em JHEP} {\bfseries 10}
  (2019) 018}, \href{http://arxiv.org/abs/1904.10940}{{\ttfamily
  arXiv:1904.10940 [hep-th]}}.

\bibitem{Chang:2021wvv}
C.-M. Chang, Y.-t. Huang, Z.-X. Huang, and W.~Li, ``{Bulk locality from the
  celestial amplitude},''
  \href{http://dx.doi.org/10.21468/SciPostPhys.12.5.176}{{\em SciPost Phys.}
  {\bfseries 12} no.~5, (2022) 176},
  \href{http://arxiv.org/abs/2106.11948}{{\ttfamily arXiv:2106.11948
  [hep-th]}}.

\bibitem{Chang:2023ttm}
C.-M. Chang, R.~Liu, and W.-J. Ma, ``{Split representation in celestial
  holography},'' \href{http://arxiv.org/abs/2311.08736}{{\ttfamily
  arXiv:2311.08736 [hep-th]}}.

\bibitem{Weinberg:1965nx}
S.~Weinberg, ``{Infrared photons and gravitons},''
  \href{http://dx.doi.org/10.1103/PhysRev.140.B516}{{\em Phys. Rev.} {\bfseries
  140} (1965) B516--B524}.

\bibitem{Arkani-Hamed:2017jhn}
N.~Arkani-Hamed, T.-C. Huang, and Y.-t. Huang, ``{Scattering amplitudes for all
  masses and spins},'' \href{http://dx.doi.org/10.1007/JHEP11(2021)070}{{\em
  JHEP} {\bfseries 11} (2021) 070},
  \href{http://arxiv.org/abs/1709.04891}{{\ttfamily arXiv:1709.04891
  [hep-th]}}.

\bibitem{Atanasov:2021cje}
A.~Atanasov, W.~Melton, A.-M. Raclariu, and A.~Strominger, ``{Conformal block
  expansion in celestial CFT},''
  \href{http://dx.doi.org/10.1103/PhysRevD.104.126033}{{\em Phys. Rev. D}
  {\bfseries 104} no.~12, (2021) 126033},
  \href{http://arxiv.org/abs/2104.13432}{{\ttfamily arXiv:2104.13432
  [hep-th]}}.

\bibitem{Camanho:2014apa}
X.~O. Camanho, J.~D. Edelstein, J.~Maldacena, and A.~Zhiboedov, ``{Causality
  Constraints on Corrections to the Graviton Three-Point Coupling},''
  \href{http://dx.doi.org/10.1007/JHEP02(2016)020}{{\em JHEP} {\bfseries 02}
  (2016) 020}, \href{http://arxiv.org/abs/1407.5597}{{\ttfamily arXiv:1407.5597
  [hep-th]}}.

\bibitem{Caron-Huot:2016icg}
S.~Caron-Huot, Z.~Komargodski, A.~Sever, and A.~Zhiboedov, ``{Strings from
  Massive Higher Spins: The Asymptotic Uniqueness of the Veneziano
  Amplitude},'' \href{http://dx.doi.org/10.1007/JHEP10(2017)026}{{\em JHEP}
  {\bfseries 10} (2017) 026}, \href{http://arxiv.org/abs/1607.04253}{{\ttfamily
  arXiv:1607.04253 [hep-th]}}.

\bibitem{Kologlu:2019bco}
M.~Kologlu, P.~Kravchuk, D.~Simmons-Duffin, and A.~Zhiboedov, ``{Shocks,
  Superconvergence, and a Stringy Equivalence Principle},''
  \href{http://dx.doi.org/10.1007/JHEP11(2020)096}{{\em JHEP} {\bfseries 11}
  (2020) 096}, \href{http://arxiv.org/abs/1904.05905}{{\ttfamily
  arXiv:1904.05905 [hep-th]}}.

\bibitem{Bern:2021ppb}
Z.~Bern, D.~Kosmopoulos, and A.~Zhiboedov, ``{Gravitational effective field
  theory islands, low-spin dominance, and the four-graviton amplitude},''
  \href{http://dx.doi.org/10.1088/1751-8121/ac0e51}{{\em J. Phys. A} {\bfseries
  54} no.~34, (2021) 344002}, \href{http://arxiv.org/abs/2103.12728}{{\ttfamily
  arXiv:2103.12728 [hep-th]}}.

\bibitem{Caron-Huot:2022ugt}
S.~Caron-Huot, Y.-Z. Li, J.~Parra-Martinez, and D.~Simmons-Duffin, ``{Causality
  constraints on corrections to Einstein gravity},''
  \href{http://dx.doi.org/10.1007/JHEP05(2023)122}{{\em JHEP} {\bfseries 05}
  (2023) 122}, \href{http://arxiv.org/abs/2201.06602}{{\ttfamily
  arXiv:2201.06602 [hep-th]}}.

\bibitem{Caron-Huot:2022jli}
S.~Caron-Huot, Y.-Z. Li, J.~Parra-Martinez, and D.~Simmons-Duffin, ``{Graviton
  partial waves and causality in higher dimensions},''
  \href{http://dx.doi.org/10.1103/PhysRevD.108.026007}{{\em Phys. Rev. D}
  {\bfseries 108} no.~2, (2023) 026007},
  \href{http://arxiv.org/abs/2205.01495}{{\ttfamily arXiv:2205.01495
  [hep-th]}}.

\bibitem{Haring:2023zwu}
K.~H\"aring and A.~Zhiboedov, ``{The Stringy S-matrix Bootstrap: Maximal Spin
  and Superpolynomial Softness},''
  \href{http://arxiv.org/abs/2311.13631}{{\ttfamily arXiv:2311.13631
  [hep-th]}}.

\bibitem{Ball:2022bgg}
A.~Ball, ``{Celestial locality and the Jacobi identity},''
  \href{http://dx.doi.org/10.1007/JHEP01(2023)146}{{\em JHEP} {\bfseries 01}
  (2023) 146}, \href{http://arxiv.org/abs/2211.09151}{{\ttfamily
  arXiv:2211.09151 [hep-th]}}.

\bibitem{Ball:2023sdz}
A.~Ball, Y.~Hu, and S.~Pasterski, ``{Multicollinear Singularities in Celestial
  CFT},'' \href{http://arxiv.org/abs/2309.16602}{{\ttfamily arXiv:2309.16602
  [hep-th]}}.

\bibitem{Guevara:2021abz}
A.~Guevara, E.~Himwich, M.~Pate, and A.~Strominger, ``{Holographic symmetry
  algebras for gauge theory and gravity},''
  \href{http://dx.doi.org/10.1007/JHEP11(2021)152}{{\em JHEP} {\bfseries 11}
  (2021) 152}, \href{http://arxiv.org/abs/2103.03961}{{\ttfamily
  arXiv:2103.03961 [hep-th]}}.

\bibitem{Strominger:2021lvk}
A.~Strominger, ``{$w_{1+\infty}$ Algebra and the Celestial Sphere: Infinite
  Towers of Soft Graviton, Photon, and Gluon Symmetries},''
  \href{http://dx.doi.org/10.1103/PhysRevLett.127.221601}{{\em Phys. Rev.
  Lett.} {\bfseries 127} no.~22, (2021) 221601},
  \href{http://arxiv.org/abs/2105.14346}{{\ttfamily arXiv:2105.14346
  [hep-th]}}.

\bibitem{Himwich:2021dau}
E.~Himwich, M.~Pate, and K.~Singh, ``{Celestial operator product expansions and
  w$_{1+\infty}$ symmetry for all spins},''
  \href{http://dx.doi.org/10.1007/JHEP01(2022)080}{{\em JHEP} {\bfseries 01}
  (2022) 080}, \href{http://arxiv.org/abs/2108.07763}{{\ttfamily
  arXiv:2108.07763 [hep-th]}}.

\bibitem{Costello:2022wso}
K.~Costello and N.~M. Paquette, ``{Celestial holography meets twisted
  holography: 4d amplitudes from chiral correlators},''
  \href{http://dx.doi.org/10.1007/JHEP10(2022)193}{{\em JHEP} {\bfseries 10}
  (2022) 193}, \href{http://arxiv.org/abs/2201.02595}{{\ttfamily
  arXiv:2201.02595 [hep-th]}}.

\bibitem{Monteiro:2022lwm}
R.~Monteiro, ``{Celestial chiral algebras, colour-kinematics duality and
  integrability},'' \href{http://dx.doi.org/10.1007/JHEP01(2023)092}{{\em JHEP}
  {\bfseries 01} (2023) 092}, \href{http://arxiv.org/abs/2208.11179}{{\ttfamily
  arXiv:2208.11179 [hep-th]}}.

\bibitem{Guevara:2022qnm}
A.~Guevara, ``{Towards Gravity From a Color Symmetry},''
  \href{http://arxiv.org/abs/2209.00696}{{\ttfamily arXiv:2209.00696
  [hep-th]}}.

\bibitem{Witten:2003nn}
E.~Witten, ``{Perturbative gauge theory as a string theory in twistor space},''
  \href{http://dx.doi.org/10.1007/s00220-004-1187-3}{{\em Commun. Math. Phys.}
  {\bfseries 252} (2004) 189--258},
  \href{http://arxiv.org/abs/hep-th/0312171}{{\ttfamily arXiv:hep-th/0312171}}.

\bibitem{Arkani-Hamed:2009hub}
N.~Arkani-Hamed, F.~Cachazo, C.~Cheung, and J.~Kaplan, ``{The S-Matrix in
  Twistor Space},'' \href{http://dx.doi.org/10.1007/JHEP03(2010)110}{{\em JHEP}
  {\bfseries 03} (2010) 110}, \href{http://arxiv.org/abs/0903.2110}{{\ttfamily
  arXiv:0903.2110 [hep-th]}}.

\bibitem{Sharma:2021gcz}
A.~Sharma, ``{Ambidextrous light transforms for celestial amplitudes},''
  \href{http://dx.doi.org/10.1007/JHEP01(2022)031}{{\em JHEP} {\bfseries 01}
  (2022) 031}, \href{http://arxiv.org/abs/2107.06250}{{\ttfamily
  arXiv:2107.06250 [hep-th]}}.

\bibitem{Jiang:2021ovh}
H.~Jiang, ``{Holographic chiral algebra: supersymmetry, infinite Ward
  identities, and EFTs},''
  \href{http://dx.doi.org/10.1007/JHEP01(2022)113}{{\em JHEP} {\bfseries 01}
  (2022) 113}, \href{http://arxiv.org/abs/2108.08799}{{\ttfamily
  arXiv:2108.08799 [hep-th]}}.

\bibitem{Jiang:2021csc}
H.~Jiang, ``{Celestial OPEs and w$_{1+\infty}$ algebra from worldsheet in
  string theory},'' \href{http://dx.doi.org/10.1007/JHEP01(2022)101}{{\em JHEP}
  {\bfseries 01} (2022) 101}, \href{http://arxiv.org/abs/2110.04255}{{\ttfamily
  arXiv:2110.04255 [hep-th]}}.

\bibitem{Adamo:2021lrv}
T.~Adamo, L.~Mason, and A.~Sharma, ``{Celestial $w_{1+\infty}$ Symmetries from
  Twistor Space},'' \href{http://dx.doi.org/10.3842/SIGMA.2022.016}{{\em SIGMA}
  {\bfseries 18} (2022) 016}, \href{http://arxiv.org/abs/2110.06066}{{\ttfamily
  arXiv:2110.06066 [hep-th]}}.

\bibitem{Ahn:2021erj}
C.~Ahn, ``{Towards a supersymmetric w$_{1+\infty}$ symmetry in the celestial
  conformal field theory},''
  \href{http://dx.doi.org/10.1103/PhysRevD.105.086028}{{\em Phys. Rev. D}
  {\bfseries 105} no.~8, (2022) 086028},
  \href{http://arxiv.org/abs/2111.04268}{{\ttfamily arXiv:2111.04268
  [hep-th]}}.

\bibitem{Ball:2021tmb}
A.~Ball, S.~A. Narayanan, J.~Salzer, and A.~Strominger, ``{Perturbatively exact
  w$_{1+\infty}$ asymptotic symmetry of quantum self-dual gravity},''
  \href{http://dx.doi.org/10.1007/JHEP01(2022)114}{{\em JHEP} {\bfseries 01}
  (2022) 114}, \href{http://arxiv.org/abs/2111.10392}{{\ttfamily
  arXiv:2111.10392 [hep-th]}}.

\bibitem{Mago:2021wje}
J.~Mago, L.~Ren, A.~Y. Srikant, and A.~Volovich, ``{Deformed $w_{1+\infty}$
  Algebras in the Celestial CFT},''
  \href{http://dx.doi.org/10.3842/SIGMA.2023.044}{{\em SIGMA} {\bfseries 19}
  (2023) 044}, \href{http://arxiv.org/abs/2111.11356}{{\ttfamily
  arXiv:2111.11356 [hep-th]}}.

\bibitem{Freidel:2021ytz}
L.~Freidel, D.~Pranzetti, and A.-M. Raclariu, ``{Higher spin dynamics in
  gravity and w1+\ensuremath{\infty} celestial symmetries},''
  \href{http://dx.doi.org/10.1103/PhysRevD.106.086013}{{\em Phys. Rev. D}
  {\bfseries 106} no.~8, (2022) 086013},
  \href{http://arxiv.org/abs/2112.15573}{{\ttfamily arXiv:2112.15573
  [hep-th]}}.

\bibitem{Ahn:2022vfw}
C.~Ahn, ``{The $ \mathcal{N} $ = 2 supersymmetric w$_{1+\infty}$ symmetry in
  the two-dimensional SYK models},''
  \href{http://dx.doi.org/10.1007/JHEP05(2022)115}{{\em JHEP} {\bfseries 05}
  (2022) 115}, \href{http://arxiv.org/abs/2203.03105}{{\ttfamily
  arXiv:2203.03105 [hep-th]}}.

\bibitem{Ahn:2022qex}
C.~Ahn, ``{N=4 supersymmetric linear
  $W\ensuremath{\infty}[\ensuremath{\lambda}]$ algebra},''
  \href{http://dx.doi.org/10.1103/PhysRevD.106.026008}{{\em Phys. Rev. D}
  {\bfseries 106} no.~2, (2022) 026008},
  \href{http://arxiv.org/abs/2205.04024}{{\ttfamily arXiv:2205.04024
  [hep-th]}}.

\bibitem{Ahn:2022orj}
C.~Ahn, ``{The structure of the $\mathcal{N}=4$ supersymmetric linear
  $W_{\infty }[\lambda ]$ algebra},''
  \href{http://dx.doi.org/10.1140/epjc/s10052-023-11752-z}{{\em Eur. Phys. J.
  C} {\bfseries 83} no.~7, (2023) 615},
  \href{http://arxiv.org/abs/2208.07000}{{\ttfamily arXiv:2208.07000
  [hep-th]}}.

\bibitem{Ren:2022sws}
L.~Ren, M.~Spradlin, A.~Yelleshpur~Srikant, and A.~Volovich, ``{On effective
  field theories with celestial duals},''
  \href{http://dx.doi.org/10.1007/JHEP08(2022)251}{{\em JHEP} {\bfseries 08}
  (2022) 251}, \href{http://arxiv.org/abs/2206.08322}{{\ttfamily
  arXiv:2206.08322 [hep-th]}}.

\bibitem{Bu:2022iak}
W.~Bu, S.~Heuveline, and D.~Skinner, ``{Moyal deformations, W$_{1+\infty}$ and
  celestial holography},''
  \href{http://dx.doi.org/10.1007/JHEP12(2022)011}{{\em JHEP} {\bfseries 12}
  (2022) 011}, \href{http://arxiv.org/abs/2208.13750}{{\ttfamily
  arXiv:2208.13750 [hep-th]}}.

\bibitem{Monteiro:2022xwq}
R.~Monteiro, ``{From Moyal deformations to chiral higher-spin theories and to
  celestial algebras},'' \href{http://dx.doi.org/10.1007/JHEP03(2023)062}{{\em
  JHEP} {\bfseries 03} (2023) 062},
  \href{http://arxiv.org/abs/2212.11266}{{\ttfamily arXiv:2212.11266
  [hep-th]}}.

\bibitem{Mason:2022hly}
L.~Mason, ``{Gravity from holomorphic discs and celestial $Lw_{1+\infty }$
  symmetries},'' \href{http://dx.doi.org/10.1007/s11005-023-01735-2}{{\em Lett.
  Math. Phys.} {\bfseries 113} no.~6, (2023) 111},
  \href{http://arxiv.org/abs/2212.10895}{{\ttfamily arXiv:2212.10895
  [hep-th]}}.

\bibitem{Banerjee:2023zip}
S.~Banerjee, H.~Kulkarni, and P.~Paul, ``{An infinite family of w$_{1+\infty}$
  invariant theories on the celestial sphere},''
  \href{http://dx.doi.org/10.1007/JHEP05(2023)063}{{\em JHEP} {\bfseries 05}
  (2023) 063}, \href{http://arxiv.org/abs/2301.13225}{{\ttfamily
  arXiv:2301.13225 [hep-th]}}.

\bibitem{Bittleston:2023bzp}
R.~Bittleston, S.~Heuveline, and D.~Skinner, ``{The celestial chiral algebra of
  self-dual gravity on Eguchi-Hanson space},''
  \href{http://dx.doi.org/10.1007/JHEP09(2023)008}{{\em JHEP} {\bfseries 09}
  (2023) 008}, \href{http://arxiv.org/abs/2305.09451}{{\ttfamily
  arXiv:2305.09451 [hep-th]}}.

\bibitem{Drozdov:2023qoy}
P.~Drozdov and T.~Kimura, ``{Structure of deformed w1+\ensuremath{\infty}
  symmetry and topological generalization in Celestial CFT},''
  \href{http://dx.doi.org/10.1016/j.physletb.2023.138272}{{\em Phys. Lett. B}
  {\bfseries 847} (2023) 138272},
  \href{http://arxiv.org/abs/2306.11693}{{\ttfamily arXiv:2306.11693
  [math-ph]}}.

\bibitem{Saha:2023abr}
A.~Saha, ``{$w_{1+\infty}$ and Carrollian Holography},''
  \href{http://arxiv.org/abs/2308.03673}{{\ttfamily arXiv:2308.03673
  [hep-th]}}.

\bibitem{Ahn:2023mdg}
C.~Ahn and M.~H. Kim, ``{The ${\cal N}=2,4$ Supersymmetric Linear
  $W_{\infty}[\lambda]$ Algebras for Generic $\lambda$ Parameter},''
  \href{http://arxiv.org/abs/2309.01537}{{\ttfamily arXiv:2309.01537
  [hep-th]}}.

\bibitem{Banerjee:2023jne}
S.~Banerjee, H.~Kulkarni, and P.~Paul, ``{Celestial OPE in Self Dual
  Gravity},'' \href{http://arxiv.org/abs/2311.06485}{{\ttfamily
  arXiv:2311.06485 [hep-th]}}.

\bibitem{Taylor:2023ajd}
T.~R. Taylor and B.~Zhu, ``{w(1+infinity) Algebra with a Cosmological Constant
  and the Celestial Sphere},''
  \href{http://arxiv.org/abs/2312.00876}{{\ttfamily arXiv:2312.00876
  [hep-th]}}.

\bibitem{Crawley:2023brz}
E.~Crawley, A.~Guevara, E.~Himwich, and A.~Strominger, ``{Self-dual black holes
  in celestial holography},''
  \href{http://dx.doi.org/10.1007/JHEP09(2023)109}{{\em JHEP} {\bfseries 09}
  (2023) 109}, \href{http://arxiv.org/abs/2302.06661}{{\ttfamily
  arXiv:2302.06661 [hep-th]}}.

\bibitem{Hamada:2018vrw}
Y.~Hamada and G.~Shiu, ``{Infinite Set of Soft Theorems in Gauge-Gravity
  Theories as Ward-Takahashi Identities},''
  \href{http://dx.doi.org/10.1103/PhysRevLett.120.201601}{{\em Phys. Rev.
  Lett.} {\bfseries 120} no.~20, (2018) 201601},
  \href{http://arxiv.org/abs/1801.05528}{{\ttfamily arXiv:1801.05528
  [hep-th]}}.

\bibitem{Li:2018gnc}
Z.-Z. Li, H.-H. Lin, and S.-Q. Zhang, ``{Infinite Soft Theorems from Gauge
  Symmetry},'' \href{http://dx.doi.org/10.1103/PhysRevD.98.045004}{{\em Phys.
  Rev. D} {\bfseries 98} no.~4, (2018) 045004},
  \href{http://arxiv.org/abs/1802.03148}{{\ttfamily arXiv:1802.03148
  [hep-th]}}.

\bibitem{Pasterski:2021dqe}
S.~Pasterski, A.~Puhm, and E.~Trevisani, ``{Revisiting the conformally soft
  sector with celestial diamonds},''
  \href{http://dx.doi.org/10.1007/JHEP11(2021)143}{{\em JHEP} {\bfseries 11}
  (2021) 143}, \href{http://arxiv.org/abs/2105.09792}{{\ttfamily
  arXiv:2105.09792 [hep-th]}}.

\bibitem{Pasterski:2021fjn}
S.~Pasterski, A.~Puhm, and E.~Trevisani, ``{Celestial diamonds: conformal
  multiplets in celestial CFT},''
  \href{http://dx.doi.org/10.1007/JHEP11(2021)072}{{\em JHEP} {\bfseries 11}
  (2021) 072}, \href{http://arxiv.org/abs/2105.03516}{{\ttfamily
  arXiv:2105.03516 [hep-th]}}.

\bibitem{Stieberger:2018onx}
S.~Stieberger and T.~R. Taylor, ``{Symmetries of Celestial Amplitudes},''
  \href{http://dx.doi.org/10.1016/j.physletb.2019.03.063}{{\em Phys. Lett. B}
  {\bfseries 793} (2019) 141--143},
  \href{http://arxiv.org/abs/1812.01080}{{\ttfamily arXiv:1812.01080
  [hep-th]}}.

\bibitem{Law:2019glh}
Y.~T.~A. Law and M.~Zlotnikov, ``{Poincar\'e constraints on celestial
  amplitudes},'' \href{http://dx.doi.org/10.1007/JHEP03(2020)085}{{\em JHEP}
  {\bfseries 03} (2020) 085}, \href{http://arxiv.org/abs/1910.04356}{{\ttfamily
  arXiv:1910.04356 [hep-th]}}. [Erratum: JHEP 04, 202 (2020)].

\bibitem{Law:2020tsg}
Y.~T.~A. Law and M.~Zlotnikov, ``{Massive Spinning Bosons on the Celestial
  Sphere},'' \href{http://dx.doi.org/10.1007/JHEP06(2020)079}{{\em JHEP}
  {\bfseries 06} (2020) 079}, \href{http://arxiv.org/abs/2004.04309}{{\ttfamily
  arXiv:2004.04309 [hep-th]}}.

\bibitem{Iacobacci:2020por}
L.~Iacobacci and W.~M\"uck, ``{Conformal Primary Basis for Dirac Spinors},''
  \href{http://dx.doi.org/10.1103/PhysRevD.102.106025}{{\em Phys. Rev. D}
  {\bfseries 102} no.~10, (2020) 106025},
  \href{http://arxiv.org/abs/2009.02938}{{\ttfamily arXiv:2009.02938
  [hep-th]}}.

\bibitem{Narayanan:2020amh}
S.~A. Narayanan, ``{Massive Celestial Fermions},''
  \href{http://dx.doi.org/10.1007/JHEP12(2020)074}{{\em JHEP} {\bfseries 12}
  (2020) 074}, \href{http://arxiv.org/abs/2009.03883}{{\ttfamily
  arXiv:2009.03883 [hep-th]}}.

\bibitem{Cachazo:2014fwa}
F.~Cachazo and A.~Strominger, ``{Evidence for a New Soft Graviton Theorem},''
  \href{http://arxiv.org/abs/1404.4091}{{\ttfamily arXiv:1404.4091 [hep-th]}}.

\bibitem{Elvang:2016qvq}
H.~Elvang, C.~R.~T. Jones, and S.~G. Naculich, ``{Soft Photon and Graviton
  Theorems in Effective Field Theory},''
  \href{http://dx.doi.org/10.1103/PhysRevLett.118.231601}{{\em Phys. Rev.
  Lett.} {\bfseries 118} no.~23, (2017) 231601},
  \href{http://arxiv.org/abs/1611.07534}{{\ttfamily arXiv:1611.07534
  [hep-th]}}.

\bibitem{Fan:2019emx}
W.~Fan, A.~Fotopoulos, and T.~R. Taylor, ``{Soft Limits of Yang-Mills
  Amplitudes and Conformal Correlators},''
  \href{http://dx.doi.org/10.1007/JHEP05(2019)121}{{\em JHEP} {\bfseries 05}
  (2019) 121}, \href{http://arxiv.org/abs/1903.01676}{{\ttfamily
  arXiv:1903.01676 [hep-th]}}.

\bibitem{Pate:2019mfs}
M.~Pate, A.-M. Raclariu, and A.~Strominger, ``{Conformally Soft Theorem in
  Gauge Theory},'' \href{http://dx.doi.org/10.1103/PhysRevD.100.085017}{{\em
  Phys. Rev. D} {\bfseries 100} no.~8, (2019) 085017},
  \href{http://arxiv.org/abs/1904.10831}{{\ttfamily arXiv:1904.10831
  [hep-th]}}.

\bibitem{Adamo:2019ipt}
T.~Adamo, L.~Mason, and A.~Sharma, ``{Celestial amplitudes and conformal soft
  theorems},'' \href{http://dx.doi.org/10.1088/1361-6382/ab42ce}{{\em Class.
  Quant. Grav.} {\bfseries 36} no.~20, (2019) 205018},
  \href{http://arxiv.org/abs/1905.09224}{{\ttfamily arXiv:1905.09224
  [hep-th]}}.

\bibitem{Puhm:2019zbl}
A.~Puhm, ``{Conformally Soft Theorem in Gravity},''
  \href{http://dx.doi.org/10.1007/JHEP09(2020)130}{{\em JHEP} {\bfseries 09}
  (2020) 130}, \href{http://arxiv.org/abs/1905.09799}{{\ttfamily
  arXiv:1905.09799 [hep-th]}}.

\bibitem{Guevara:2019ypd}
A.~Guevara, ``{Notes on Conformal Soft Theorems and Recursion Relations in
  Gravity},'' \href{http://arxiv.org/abs/1906.07810}{{\ttfamily
  arXiv:1906.07810 [hep-th]}}.

\bibitem{Bautista:2019tdr}
Y.~F. Bautista and A.~Guevara, ``{From Scattering Amplitudes to Classical
  Physics: Universality, Double Copy and Soft Theorems},''
  \href{http://arxiv.org/abs/1903.12419}{{\ttfamily arXiv:1903.12419
  [hep-th]}}.

\bibitem{Banerjee:2019aoy}
S.~Banerjee, P.~Pandey, and P.~Paul, ``{Conformal properties of soft operators:
  Use of null states},''
  \href{http://dx.doi.org/10.1103/PhysRevD.101.106014}{{\em Phys. Rev. D}
  {\bfseries 101} no.~10, (2020) 106014},
  \href{http://arxiv.org/abs/1902.02309}{{\ttfamily arXiv:1902.02309
  [hep-th]}}.

\bibitem{Banerjee:2019tam}
S.~Banerjee and P.~Pandey, ``{Conformal properties of soft-operators. Part II.
  Use of null-states},'' \href{http://dx.doi.org/10.1007/JHEP02(2020)067}{{\em
  JHEP} {\bfseries 02} (2020) 067},
  \href{http://arxiv.org/abs/1906.01650}{{\ttfamily arXiv:1906.01650
  [hep-th]}}.

\bibitem{Banerjee:2020kaa}
S.~Banerjee, S.~Ghosh, and R.~Gonzo, ``{BMS symmetry of celestial OPE},''
  \href{http://dx.doi.org/10.1007/JHEP04(2020)130}{{\em JHEP} {\bfseries 04}
  (2020) 130}, \href{http://arxiv.org/abs/2002.00975}{{\ttfamily
  arXiv:2002.00975 [hep-th]}}.

\bibitem{Banerjee:2020zlg}
S.~Banerjee, S.~Ghosh, and P.~Paul, ``{MHV Graviton Scattering Amplitudes and
  Current Algebra on the Celestial Sphere},''
  \href{http://arxiv.org/abs/2008.04330}{{\ttfamily arXiv:2008.04330
  [hep-th]}}.

\bibitem{Banerjee:2021cly}
S.~Banerjee, S.~Ghosh, and S.~Satyam~Samal, ``{Subsubleading soft graviton
  symmetry and MHV graviton scattering amplitudes},''
  \href{http://arxiv.org/abs/2104.02546}{{\ttfamily arXiv:2104.02546
  [hep-th]}}.

\bibitem{Banerjee:2021dlm}
S.~Banerjee, S.~Ghosh, and P.~Paul, ``{(Chiral) Virasoro invariance of the
  tree-level MHV graviton scattering amplitudes},''
  \href{http://arxiv.org/abs/2108.04262}{{\ttfamily arXiv:2108.04262
  [hep-th]}}.

\bibitem{Banerjee:2020vnt}
S.~Banerjee and S.~Ghosh, ``{MHV Gluon Scattering Amplitudes from Celestial
  Current Algebras},'' \href{http://arxiv.org/abs/2011.00017}{{\ttfamily
  arXiv:2011.00017 [hep-th]}}.

\bibitem{Simmons-Duffin:2012juh}
D.~Simmons-Duffin, ``{Projectors, Shadows, and Conformal Blocks},''
  \href{http://dx.doi.org/10.1007/JHEP04(2014)146}{{\em JHEP} {\bfseries 04}
  (2014) 146}, \href{http://arxiv.org/abs/1204.3894}{{\ttfamily arXiv:1204.3894
  [hep-th]}}.

\bibitem{MacDowell1972}
W.~W. {MacDowell} and R.~{Roskies}, ``{Reduction of the Poincar{\'e} Group with
  Respect to the Lorentz Group},''
  \href{http://dx.doi.org/10.1063/1.1665882}{{\em Journal of Mathematical
  Physics} {\bfseries 13} no.~10, (Oct., 1972) 1585--1592}.

\end{thebibliography}\endgroup
\bibliographystyle{utphys}

\end{document}